\documentclass[aps,twocolumn,floats,showpacs,amsmath,superscriptaddress,prb]{revtex4-1}
\usepackage{graphicx}
\usepackage{dcolumn}
\usepackage{bm}
\usepackage{amsmath}
\usepackage{amsfonts}
\usepackage{amssymb}
\usepackage{xcolor}
\usepackage{soul}
\usepackage{epstopdf}
\usepackage{epsfig}
\usepackage{CJK}
\usepackage{hyperref}
\usepackage{epsfig}
\usepackage{subfigure}
\usepackage{mathrsfs}
\usepackage{cases}
\usepackage{bbm}

\begin{document}


\title{Instability caused by the fermion-fermion interactions combined with rotational and
particle-hole asymmetries in three-dimensional materials with quadratic band touching}

\date{\today}

\author{Jing Wang}
\altaffiliation{E-mail address: jing$\textunderscore$wang@tju.edu.cn}
\affiliation{Department of Physics, Tianjin University, Tianjin 300072, P.R. China}
\affiliation{Department of Modern Physics, University of Science and
Technology of China, Hefei, Anhui 230026, P.R. China}

\begin{abstract}
We investigate the role of four-fermion interactions, rotational and
particle-hole asymmetries, and their interplay in three-dimensional systems
with a quadratic band touching point by virtue of the renormalization group
approach, which allows to treat all these facets unbiasedly. The coupled flow evolutions
of interaction parameters are derived by taking into account one-loop
corrections in order to explore the behaviors of low-energy states. We find
four-fermion interaction can drive Gaussian fixed points to
be unstable in the low-energy regime. In addition, the rotational and particle-hole asymmetries,
together with the fermion-fermion interactions conspire to split the trajectories
of distinct types of fermionic couplings and induce superconductivity instability with appropriate
starting conditions.  Furthermore, we present the schematic phase diagrams in the parameter space,
showing the overall behaviors of states in the low-energy regime caused by both fermionic interactions
and asymmetries.
\end{abstract}


\maketitle


\section{Introduction}

Systems with a quadratic band touching (QBT) point in their electronic structures recently
have become a much studied subject in condensed matter physics~\cite{Chong2008PRB,
Fradkin2008PRB,Fradkin2009PRL,Vafek2012PRB,Vafek2014PRB,Herbut2012PRB,Herbut2014PRL,
Herbut2014PRB,Herbut2015PRB,Murray2015PRB,Herbut2016PRB,Herbut2017PRB,Herbut2017PRB_2,
Nandkishore2017PRB,Nandkishore2018PRB,Nandkishore2018PRB_2,Mandal2018AP}.
In a sharp contrast to the conventional Fermi metals owning a finite Fermi
surface~\cite{Altland2006Book} or Dirac/Weyl semimetals possessing
several discrete Dirac points and linear energy dispersions~\cite{Neto2009RMP},
the two-dimensional (2D) QBT systems have a QBT point at which the density of
states is finite~\cite{Fradkin2009PRL,Vafek2014PRB} such as the bilayer graphene~\cite{Nilsson2008PRB,Nandkishore2010PRB,Vafek2012PRB,
MacDonald2011PRL,Lemonik2012PRB,MacDonald2012PRB}. This unconventional structure
of Fermi surface can bring a range of interesting phenomena, for instance the quantum anomalous
Hall (QAH) effect~\cite{Sinova2004PRL,Murakami2003Science,Hirsch1999PRL} and
quantum spin Hall (QSH) effect~\cite{Kane2005PRL,Bernevig2006PRL,Bernevig2006Science}.
Recently, it has been found that both QAH and QSH can be stabilized in the checkerboard
lattice model with a 2D QBT system by the short-ranged four-fermion interactions~\cite{Fradkin2009PRL,
Vafek2014PRB}.  Several interesting behaviors and instabilities of these topological
insulators in the presence of distinct sorts of impurities have also been examined
and addressed in Ref.~\cite{Wang2017}.  Besides the 2D QBT systems~\cite{Chong2008PRB,
Fradkin2008PRB,Fradkin2009PRL,Vafek2012PRB,Vafek2014PRB,Herbut2012PRB,Mandal1808}, three-dimensional
(3D) QBT electronic systems~\cite{Luttinger1956PR,Nandkishore2017PRB,Nandkishore2018PRB,Nandkishore2018PRB_2,
Tsidilkovski1997Book, Murakami2004PRB,Moon2013PRL,Witczak2014ARCMP,Kondo2015NatC,Herbut2015PRB,Murray2015PRB,
Herbut2016PRB,Herbut2017PRB,Herbut2017PRB_2,Savary2014PRX,Savary2014PRB,Moon1811,
Lai2014arXiv,Goswami2017PRB,Szabo2018arXiv,Foster2019PRB},
such as Luttinger semimetals~\cite{Luttinger1956PR,
Murakami2004PRB,Lai2014arXiv,Goswami2017PRB,Szabo2018arXiv,Foster2019PRB}, have also attracted much attention due to their unconventional 3D
dispersions of low-energy excitations and fermion-fermion interactions~\cite{Luttinger1956PR,
Herbut2012PRB,Herbut2014PRL,Herbut2014PRB,Herbut2015PRB,Herbut2016PRB,Herbut2017PRB,Herbut2017PRB_2,
Nandkishore2017PRB,Nandkishore2018PRB,Nandkishore2018PRB_2,
Lai2014arXiv,Goswami2017PRB,Szabo2018arXiv,Foster2019PRB}.

In 3D QBT systems, such as gray tin, HgTe~\cite{Tsidilkovski1997Book} or Luttinger semimetals~\cite{Luttinger1956PR,Murakami2004PRB}, the electron-electron interactions
may result in non-Fermi liquid behaviors~\cite{Herbut2014PRB,Herbut2014PRL,
Vafek2014PRB,Herbut2015PRB,Herbut2016PRB,Herbut2006PRL}. Additionally, the phase
transitions/potential quantum phase transitions might be accompanied by lots of
singular and interesting quantum critical behaviors in the vicinity of
corresponding quantum critical point (QCP) at the low-energy
regime~\cite{Herbut2014PRL,Herbut2015PRB,Herbut2016PRB}, which are closely related to
the unconventional structure of Fermi surface and low-energy energy dispersions in 3D
QBT systems. In order to investigate the quantum criticality and unusually physical behaviors
in 3D QBT systems, it is worth exploring whether the system harbors any fixed points
and judging which fixed points are stable or unstable in the presence of different types of
electron-electron interactions and rotational/particle-hole asymmetries in the low-energy
regime. Fortunately, one usually can employ the powerful renormalization group (RG) approach~\cite{Wilson1975RMP,Polchinski9210046,Shankar1994RMP,Herbut2007Book} to seek
the potential fixed points at the lowest-energy limit by treating distinct sorts of
interactions on the same footing. After determining the potential (rescaled)
fixed points~\cite{Vafek2012PRB,Vafek2014PRB} and stable fixed points in the low-energy
regime, we can explore the associated physical behaviors in the low-energy regime
and investigate the corresponding phase transitions/quantum phase transitions in the
vicinity of these fixed points.

In this work, we study how the four-fermion interactions,
rotational and particle-hole asymmetries, and their interplay influence  the
low-energy states of 3D QBT semimetal, which is a typically 3D electronic system with
quadratic band touching~\cite{Herbut2012PRB,Herbut2014PRL,Herbut2015PRB,Herbut2016PRB,Savary2014PRX,Savary2014PRB,Moon1811}.
To this end, we adopt the RG framework~\cite{Wilson1975RMP,Polchinski9210046,Shankar1994RMP,
Herbut2007Book} to unbiasedly treat all possible effects and contributions.
To be specific, the rotational and particle-hole asymmetries
as well as all six short-ranged fermion-fermion interactions are considered at
the same footing to derive the coupled flow equations of the interaction parameters.
Based on these, we subsequently study the behaviors of low-energy states. Concretely,
we, at the outset, only switch on the four-fermion interactions with preservation of
rotational and particle-hole symmetries. After completing attentive RG analysis and
numerically dealing with the associated coupled flow equations, we find that the
short-ranged fermionic interaction can drive noninteracting Gaussian fixed point to become
unstable and eventually flow to the strong coupling in the low-energy regime as long as the
initial values of these coupling are adequately large. Thereafter, the rotational and
particle-hole asymmetries are also turned on. On one hand, the trajectories of
distinct types of four-fermion couplings that are degenerated in the noninteracting
situation are unambiguously split by the interplay
between four-fermion interactions and asymmetries.
On the other, we find there exists a relatively fixed point or critical point at certain
critical energy scale as long as the starting conditions are satisfied. Superconductivity
instability is always triggered and conventionally
via approaching this fixed point. In order to obviously exhibit
the overall behaviors of states in the low-energy regime, the
schematic phase diagrams are provided upon varying the particle-hole
and rotational asymmetric parameters $\alpha$ and $\delta$.

The rest paper is organized as follows. In Sec.~\ref{Sec_model}, we provide our microscopic
model and address the effective theory for 3D QBT semimetals.  The Sec.~\ref{Sec_coupled_equations}
is accompanied to complete the one-loop RG analysis and derive the coupled flow equations of
all fermion-fermion couplings. In Sec.~\ref{Sec_interaction}, we numerically evaluate the correlated
coupled equations and discuss the influence of four-fermion interaction on the low-energy states
of 3D QBT systems. We next try to inspect the effects of low-energy states by incorporating
into the contributions from both the six types of short-ranged fermionic interactions and rotational
and particle-hole asymmetries in Sec.~\ref{Sec_interaction_asymm}.
We finally present a short summary in Sec.~\ref{Sec_summary}.

\begin{figure}
\centering
\includegraphics[width=3.0in]{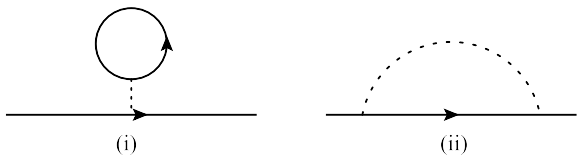}
\vspace{-0.15cm}
\caption{One-loop corrections to the fermion propagator (the dashed
line indicates the interactions).}\label{Fig_fermion_propagator_correction}
\end{figure}

\section{Model and Effective theory}\label{Sec_model}

Within this work, our focus is on the the following standard 3D QBT
semimetals~\cite{Luttinger1956PR,Herbut2016PRB}, whose low-energy quasiparticle excitations
and their interactions can be described by
\begin{eqnarray}
\mathcal{L}\!=\!\psi^\dagger(\tau,\mathbf{x})(\partial_\tau\!+\!H_0)\psi(\tau,\mathbf{x})
\!+\!\!\sum^5_{i=0}g_i[\psi^\dagger(\tau,\mathbf{x})\gamma_i\psi(\tau,\mathbf{x})]^2,\label{Eq_model}
\end{eqnarray}
with $\tau$ and $\psi(\tau,\mathbf{x})$ respectively representing the imaginary time and
four-component Grassmann field $\psi(\tau,\mathbf{x})=(\psi_1,\psi_2,\psi_3,\psi_4)^T$.
Here, the $\gamma$ matrices, which stem from one of the (two possible) irreducible four-dimensional
Hermitian representations~\cite{Herbut2015PRB,Herbut2016PRB}, own five components and satisfy
the standard Clifford algebra $\{\gamma_a,\gamma_b\}=2\delta_{ab}$, $a,b=1-5$. In addition,
$\gamma_0$ serves as the identity matrix.

We firstly consider the free terms of the low-energy quasiparticles, which can be expressed as
\begin{eqnarray}
\mathcal{H}_0(\mathbf{p})=\alpha\mathbf{p}^2+\sum^5_{a=1}d_a(\mathbf{p})\gamma_a
+\delta\sum^5_{a=1}s_ad_a(\mathbf{p})\gamma_a,\label{Eq_H_0}
\end{eqnarray}
The parameters $\alpha$ and $\delta$ are directly linked to
the Luttinger parameters~\cite{Herbut2016PRB}. It is worth pointing
out that the particle-hole and rotational
symmetries/asymmetries of systems are closely associated with the values of
parameters $\alpha$ and $\delta$, respectively. To be specific,
the physical systems harbor both particle-hole and rotational symmetries~\cite{Herbut2012PRB,
Herbut2015PRB,Herbut2016PRB} once both of two parameters $\alpha$ and $\delta$
are tuned to $1$ and $0$, namely $\alpha_s\equiv1$ and $\delta_s\equiv0$, respectively.
On the contrary, as long as the symmetric values of
$\alpha$ and/or $\delta$ are deviated,
particle-hole and/or rotational asymmetries are
developed simultaneously. Here, the $d_a(\mathbf{p})$ factor possesses a general form in $d$
dimension~\cite{Herbut2015PRB,Herbut2016PRB}
\begin{eqnarray}
d_a(\mathbf{p})=\sqrt{\frac{d}{2(d-1)}}p_i\Lambda^a_{ij}p_j,
\end{eqnarray}
with $\Lambda^a_{ij}$ representing a $d\times d$ Gell-Mann matrices~\cite{Herbut2015PRB,Herbut2016PRB}.
Utilizing the same conventions in Ref.~\cite{Herbut2016PRB}, we choose $s_a=1$ for the
off-diagonal (indices 2,3,4) and $s_a =-1$ for the diagonal (indices 1,5) matrices.
Concretely, the five functions $d_a(\mathbf{p})$ in Eq.~(\ref{Eq_H_0}) can be
designated for our $d=3$ model by exploiting the real $l=2$ spherical
harmonics as~\cite{Herbut2012PRB,Herbut2015PRB,Herbut2016PRB}
\begin{eqnarray}
d_1(\mathbf{p})&=&\frac{\sqrt{3}}{2}(p^2_x-p^2_y),\nonumber\\
d_2(\mathbf{p})&=&\sqrt{3}p_xp_y,\nonumber\\
d_3(\mathbf{p})&=&\sqrt{3}p_xp_z,\label{Eq_d_a}\\
d_4(\mathbf{p})&=&\sqrt{3}p_yp_z,\nonumber\\
d_5(\mathbf{p})&=&\frac{1}{2}(2p^2_z-p^2_x-p^2_y).\nonumber
\end{eqnarray}

In addition, the last term of Eq.~(\ref{Eq_model}) captures all potential four-fermion
interactions of the low-energy quasiparticles, which will be focused
in Sec.~\ref{Sec_coupled_equations}. The one-loop RG analysis of our model~(\ref{Eq_model})
will be practiced in forthcoming two sections for both symmetric and asymmetric cases.

\section{RG analysis and coupled flow equations}\label{Sec_coupled_equations}

Before performing a standard RG analysis, it is convenient to rescale the
momenta and energy by $\Lambda_0$ which is tied to the lattice constant,
i.e. $\mathbf{p}\rightarrow\mathbf{p}/\Lambda_0$ and  $\omega\rightarrow\omega=
\omega/\Lambda^2_0$, and designate the energy scale as $\Lambda=\Lambda_0/b$ with
$b=e^{-l}$ and $l\geq0$~\cite{Wilson1975RMP,Polchinski9210046,Shankar1994RMP,Huh2008PRB,
She2010PRB,Wang2013NJP,Herbut2007Book,She2015PRB,Wang2014PRD,Wang2015PRB,Wang2017PRB,
Wang2011PRB,Wang2013PRB,Wang2015PLA,Wang2018JPCM,Wang1806}
to represent the evolution of energy scales. Next, we are going to extract the
contributions from these one-loop diagrams and derive the coupled flow equations
for the interaction parameters.

Before moving further, it is of very importance to emphasize that
the RG evolutions of interaction parameters are closely associated with the free fixed point, which
determines the basic RG rescalings of energies, momenta, and fields.
In this work, we make $S_0\sim i\omega+\sum^5_{a=1}d_a(\mathbf{p})\gamma_a$
invariant under RG transformation as the free fixed point and correspondingly
the rescalings are
\begin{eqnarray}
\omega&\rightarrow&\omega'e^{-2l},\\
p_{i}&\rightarrow& p'_{i}e^{-l},\\
\psi(\omega,\mathbf{p})&\rightarrow&\psi'(\omega',\mathbf{p'})e^{7/2l}.
\end{eqnarray}
According to these points, the parameters $\alpha$ and $\delta$ are two constants
at the tree level. Further, as the one-loop corrections from four-fermion interactions
do not contribute to the self-energy, one can easily figure out that they both are still
energy-independent at least to one-loop level. As a consequence, we within this work regard these two
parameters as two constants that are assumed to be linked to
distinct physical situations among many others.

\subsection{In the presence of the rotational and particle-hole symmetries}\label{Sec_rot_par}

We firstly go to investigate the system in the presence of both the rotational and
particle-hole symmetries. In such circumstances, the free term $H_0$ in Eq.~(\ref{Eq_H_0})
is reduced to the compact form
\begin{eqnarray}
\mathcal{H}_0(\mathbf{p})=\sum^5_{a=1}d_a(\mathbf{p})\gamma_a,
\end{eqnarray}
where $d_a(\mathbf{p})\equiv \mathbf{p}^2\tilde{d}_a(\Omega)$ with $a=1-5$
has already been introduced in Eq.~(\ref{Eq_d_a}).
This defines the real hyperspherical harmonics $\tilde{d}_a(\Omega)$ for angular
momentum of two in general dimension, with $\Omega$ denoting the spherical
angles on the $(d-1)$ sphere in $\mathbf{p}$ space~\cite{Herbut2012PRB,
Herbut2015PRB,Herbut2016PRB}. After performing corresponding Fourier transformations
by adopting our starting point (\ref{Eq_model}), we obtain the revised
effective action
\begin{widetext}
\begin{eqnarray}
S&=&\int^{+\infty}_{-\infty}\frac{d\omega}{2\pi}\int^{\Lambda}\frac{d^d\mathbf{p}}{(2\pi)^d}
\psi^\dagger(\omega,\mathbf{p})\left[i\omega+\sum^5_{a=1}d_a(\mathbf{p})\gamma_a\right]\psi(\omega,\mathbf{p})
+\sum^5_{i=0}g_i\int^{+\infty}_{-\infty}\frac{d\omega_1d\omega_2d\omega_3}{(2\pi)^3}\int^{\Lambda}
\frac{d^d\mathbf{p}_1d^d\mathbf{p}_2d^d\mathbf{p}_3}{(2\pi)^{3d}}\nonumber\\
&&\times\psi^\dagger(\omega_1,\mathbf{p}_1)
\gamma_i\psi(\omega_2,\mathbf{p}_2)\psi^\dagger(\omega_3,\mathbf{p}_3)\gamma_i
\psi(\omega_1+\omega_2-\omega_3,\mathbf{p}_1+\mathbf{p}_2-\mathbf{p}_3),\label{Eq_eff_action}
\end{eqnarray}
\end{widetext}
which straightforwardly gives rise to the free fermion
propagator~\cite{Herbut2015PRB,Herbut2016PRB},
\begin{eqnarray}
G_0(\omega,\mathbf{p})
&=&\frac{-i\omega+\sum^5_{a=1}d_a(\mathbf{p})\gamma_a}{\omega^2+\mathbf{p}^4}.\label{Eq_propagators}
\end{eqnarray}

\begin{figure}
\centering
\includegraphics[width=3.3in]{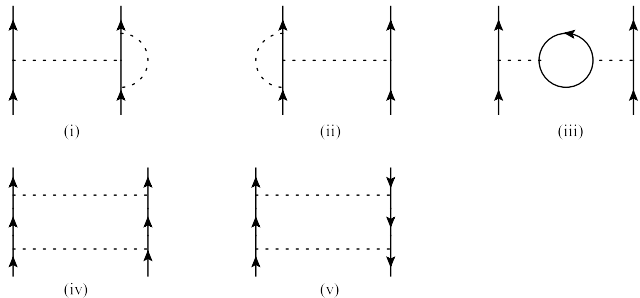}
\vspace{-0.10cm}
\caption{One-loop corrections to the fermion interacting couplings (the dashed
line indicates the interactions).}\label{Fig_fermion_interaction_correction}
\end{figure}

Subsequently, we derive the coupled flow equations of the parameters after
considering the one-loop corrections in Fig.~\ref{Fig_fermion_propagator_correction}
and Fig.~\ref{Fig_fermion_interaction_correction} and carrying out the standard RG
analysis~\cite{Shankar1994RMP,Huh2008PRB,She2010PRB,Wang2013NJP,Herbut2007Book,
She2015PRB,Wang2014PRD,Wang2015PRB,Wang2017PRB},
\begin{eqnarray}
\frac{d g_0}{dl}&=&-g_0-\frac{g_0\mathcal{F}_0}{5\pi^2},\label{Eq_flow_Eq_g_0_xi_0}\\
\frac{d g_a}{dl}&=&-g_a+\frac{4g_a(g_0-\mathcal{F}_0)-\mathcal{F}_1}{10\pi^2}.\label{Eq_flow_Eq_g_5_xi_0}
\end{eqnarray}
Here, the index $a$ runs from $1$ to $5$, namely the flows of interaction parameters $g_1-g_5$ share
with the similar structure. In order to simplify our equations, we also
assign two new parameters $\mathcal{F}_0\equiv \sum^5_{a=1}g_a$ and
$\mathcal{F}_1\equiv \sum^5_{i=0} g^2_i$. Before moving further,
we would like to address several comments on above RG equations.
On one hand, the first terms are linear, which come from the tree-level corrections of fermion-
fermion interactions as provided in Eq.~(\ref{Eq_eff_action}). On the other hand, as manifestly
depicted in Fig.~\ref{Fig_fermion_interaction_correction}, each one-loop diagram
involves two vertexes, which hence carries two sorts of fermion-fermion strengths.
This indicates that one-loop contributions, namely the second terms RG equations,
to $g_i$ with $i=0-5$ are quadratic. Although $\mathcal{F}_0$ and $\mathcal{F}_1$
own distinct structures, they, together with $g_0$ and $g_i$, ensure that the second terms'
numerators of both Eqs.~(\ref{Eq_flow_Eq_g_0_xi_0}) and (\ref{Eq_flow_Eq_g_5_xi_0})
are quadratic. Accordingly, these are well consistent with RG analysis.

\subsection{In the absence of the rotational and particle-hole symmetries}\label{Sec_non_rot_par}

We then move to the asymmetric situations. In the absence of the
rotational and particle-hole symmetries, we have the general
effective free Hamiltonian as given in Eq.~(\ref{Eq_H_0})
with the restricted conditions $\delta\neq\delta_s$ and $\alpha\neq\alpha_s$~\cite{Herbut2012PRB,
Herbut2015PRB,Herbut2016PRB}. The free fermion propagator can be
correspondingly extracted as~\cite{Herbut2015PRB,Herbut2016PRB},
\begin{eqnarray}
G_0(\omega,\mathbf{p})
\!\!&=&\!\!\frac{1}{i\omega+\alpha\mathbf{p}^2+\sum^5_{a=1}(1+\delta s_a)d_a(\mathbf{p})\gamma_a},\label{Eq_propagators_2}
\end{eqnarray}
with the asymmetric parameters $\alpha\neq\alpha_s$ and $\delta\neq\delta_s$. In order to simplify
further expressions and calculations, we here derive and designate some useful identities
by employing Eq.~(\ref{Eq_d_a}),
\begin{eqnarray}
\sum^5_{a=1}\left[d_a(\mathbf{p})\gamma_a\right]^2=(p^2_z+p^2_x+p^2_y)^2=\mathbf{p}^4,
\end{eqnarray}
and
\begin{eqnarray}
&&\sum^5_{a=1}(1+\delta s_a)^2\left[d_a(\mathbf{p})\gamma_a\right]^2\nonumber\\
&=&(1-\delta)^2(p^4_x+p^4_y+p^4_z)+2(1+4\delta+\delta^2)\nonumber\\
&&\times(p^2_xp^2_y+p^2_xp^2_z+p^2_yp^2_z)\nonumber\\
&\equiv&\mathbf{p}^4_\delta.
\end{eqnarray}

To proceed, we can derive the coupled flow equations for general values of
$\alpha$ and $\delta$, which result in the rotational and particle-hole
asymmetries via paralleling and performing the similar
procedures of the rotational and particle-hole symmetry case as shown in Sec.~\ref{Sec_rot_par},
\begin{widetext}
\begin{eqnarray}
\frac{d g_0}{dl}
&=&-g_0-\frac{1}{16\pi^3}\Bigl[g_0(\mathcal{F}_0-g_0)\mathcal{N}_{+++++}
+g_0(g_1+g_5)\mathcal{M}_3+g_0(g_2+g_3+g_4)\mathcal{M}_2
+\frac{1}{2}\mathcal{F}_1\mathcal{M}_1\Bigr],\label{Eq_flow_Eq_g_0}\\
\frac{d g_1}{dl}
&=&-g_1-\frac{1}{16\pi^3}\Bigl[-2g^2_1 \mathcal{N}_{+----}+g_1(g_0+2g_1-\mathcal{F}_0)
\mathcal{N}_{++++-}+g_1[g_0\mathcal{M}_1+g_5\mathcal{M}_5\nonumber\\
&&+(g_2+g_3+g_4)\mathcal{M}_4]+\frac{1}{2}\mathcal{F}_1
\mathcal{M}_3\Bigr],\\
\frac{d g_2}{dl}
&=&-g_2-\frac{1}{16\pi^3}\Bigl[g_2(g_0-\mathcal{F}_0)
\mathcal{N}_{-+---}+g_2[g_0\mathcal{M}_1+(g_1+g_5)\mathcal{M}_5+(g_3+g_4)\mathcal{M}_4]
+\frac{1}{2}\mathcal{F}_1
\mathcal{M}_2\Bigr],\\
\frac{d g_3}{dl}
&=&-g_3-\frac{1}{16\pi^3}\Bigl[
g_3(g_0-\mathcal{F}_0)\mathcal{N}_{--+--}+g_3[g_0\mathcal{M}_1+(g_1+g_5)\mathcal{M}_5
+(g_2+g_4)\mathcal{M}_4]+\frac{1}{2}\mathcal{F}_1
\mathcal{M}_2\Bigr],\\
\frac{d g_4}{dl}
&=&-g_4-\frac{1}{16\pi^3}\Bigl[
g_4(g_0-\mathcal{F}_0)\mathcal{N}_{---+-}+g_4[g_0\mathcal{M}_1+(g_1+g_5)\mathcal{M}_5
+(g_2+g_3)\mathcal{M}_4]+\frac{1}{2}\mathcal{F}_1
\mathcal{M}_2\Bigr],\\
\frac{d g_5}{dl}
&=&-g_5-\frac{1}{16\pi^3}\Bigl[
g_5(g_0-\mathcal{F}_0)\mathcal{N}_{----+}+g_5[g_0\mathcal{M}_1+g_1\mathcal{M}_5
+(g_2+g_3+g_4)\mathcal{M}_4]+\frac{1}{2}\mathcal{F}_1
\mathcal{M}_3\Bigr].\label{Eq_flow_Eq_g_5}
\end{eqnarray}
\end{widetext}
Again, the first and second terms of above RG equations that collect tree-level and
one-loop level corrections of fermion-fermion interactions are linear and quadratic,
respectively. Please refer to the paragraph below Eq.~(\ref{Eq_flow_Eq_g_5_xi_0}) for
detailed information. In order to write the coupled equations as the compact forms, we here bring out
several new coefficients in above coupled running Eqs.~(\ref{Eq_flow_Eq_g_0})-(\ref{Eq_flow_Eq_g_5}).
The defined parameters/functions $\mathcal{F}_0$ and $\mathcal{F}_1$ have already been provided in
Eqs.~\ref{Eq_flow_Eq_g_0_xi_0} and \ref{Eq_flow_Eq_g_5_xi_0} and $\mathcal{M}_{i}$ with $i=1-5$
are nominated as follows,
\begin{eqnarray}
\mathcal{M}_1\!\!\!&=&\!\!\!\!\int_{x,\theta,\varphi}\frac{64\sin^2\theta[\alpha^2(\alpha^2-f_\delta+x^2)]}
{[(\alpha^2-f_\delta-x^2)^2+4x^2\alpha^2]^2} ,\\
\mathcal{M}_2\!\!\!&=&\!\!\!\!\int_{x,\theta,\varphi}\frac{64\sin^2\theta
[(\alpha^2-f_\delta-x^2)^2](1+\delta)^2f_{d^2_C}}
{[(\alpha^2-f_\delta-x^2)^2+4x^2\alpha^2]^2},\\
\mathcal{M}_3\!\!\!&=&\!\!\!\!\int_{x,\theta,\varphi}\frac{64\sin^2\theta
[(\alpha^2-f_\delta-x^2)^2](1-\delta)^2f_{d^2_C}}
{[(\alpha^2-f_\delta-x^2)^2+4x^2\alpha^2]^2},\\
\mathcal{M}_4\!\!\!&=&\!\!\!\!\int_{x,\theta,\varphi}\frac{\sin^2\theta
\Big\{32[-8\alpha^2x^2](1+\delta)^2f_{d^2_C}\Bigr\}}
{[(\alpha^2-f_\delta-x^2)^2+4x^2\alpha^2]^2},\\
\mathcal{M}_5\!\!\!&=&\!\!\!\!\int_{x,\theta,\varphi}\frac{\sin^2\theta
\Big\{32[-8\alpha^2x^2](1-\delta)^2f_{d^2_C}\Bigr\}}
{[(\alpha^2-f_\delta-x^2)^2+4x^2\alpha^2]^2},
\end{eqnarray}
with introducing
\begin{eqnarray}
\int_{x,\theta,\varphi}\equiv\int^{+\infty}_{-\infty}\frac{dx}{(2\pi)}
\int^\pi_0d\theta\int^{2\pi}_0d\varphi.
\end{eqnarray}
Here the coefficients for $f_\delta$
and $f_{d^2_a}$ with $a=1-5$ are given by
\begin{eqnarray}
f_\delta\!\!\!&\equiv&\!\!\![2(1+4\delta+\delta^2)(\sin^4\theta\cos^2\varphi\sin^2\varphi
+\sin^2\theta\cos^2\theta)\nonumber\\
\!\!\!&&\!\!\!\!\!\!\!\!\!+(1-\delta)^2(\sin^4\theta\cos^4\varphi
+\sin^4\theta\sin^4\varphi+\cos^4\theta)],\\
f_{d^2_1}\!\!\!&\equiv&\!\!\!3(\sin^2\theta\cos^2\varphi-\sin^2\theta\sin^2\varphi)^2/4,\\
f_{d^2_2}\!\!\!&\equiv&\!\!\!3\sin^4\theta\sin^2\varphi\cos^2\varphi,\\
f_{d^2_3}\!\!\!&\equiv&\!\!\!3\sin^2\theta\cos^2\theta\cos^2\varphi,\\
f_{d^2_4}\!\!\!&\equiv&\!\!\!3\sin^2\theta\cos^2\theta\sin^2\varphi,\\
f_{d^2_5}\!\!\!&\equiv&\!\!\!(2\cos^2\theta-\sin^2\theta)^2/4.
\end{eqnarray}
In addition,  the coefficients $\mathcal{N}(\xi_1\xi_2\xi_3\xi_4\xi_5)$ are designated as
\begin{eqnarray}
\mathcal{N}(\xi_1\xi_2\xi_3\xi_4\xi_5)
\!\!\!&=&\!\!\!\!\!\int_{x,\theta,\varphi}
\frac{\sin^2\theta}{[(\alpha^2-f_\delta
-x^2)^2+4x^2\alpha^2]^2}\nonumber\\
&&\!\!\!\!\!\!\!\!\!\!\!\!\!\!\!\!\!\!\!\!\!\!\!\!\!\!\!\!\!\!\!\!\!\!\!\!\!
\!\!\!\!\!\!\!\!\!\!\!\!\!\!\!\!\!\!\!\!\times\Bigl\{\!\!4[-x^2(\alpha^2+f_\delta+x^2)^2
+\alpha^2(\alpha^2-f_\delta-x^2)^2
+4\alpha^2x^2\nonumber\\
&&\!\!\!\!\!\!\!\!\!\!\!\!\!\!\!\!\!\!\!\!\!\!\!\!\!\!\!\!\!\!\!\!\!\!\!\!\!
\!\!\!\!\!\!\!\!\!\!\!\!\!\!\!\!\!\!\!\!\times(\alpha^2-f_\delta)]
+4[(\alpha^2-f_\delta-x^2)^2-4\alpha^2x^2]
[3(1+\delta)^2\nonumber\\
&&\!\!\!\!\!\!\!\!\!\!\!\!\!\!\!\!\!\!\!\!\!\!\!\!\!\!\!\!\!\!\!\!\!\!\!\!\!
\!\!\!\!\!\!\!\!\!\!\!\!\!\!\!\!\!\!\!\!\times
(\xi_2f_{d^2_2}+\xi_3f_{d^2_3}+\xi_4f_{d^2_4})
+2(1-\delta)^2(\xi_1f_{d^2_1}+\xi_5f_{d^2_5})]\!\!\Bigr\}.
\end{eqnarray}
Hereby, we highlight that the five signs $\xi_i=\pm$ corresponds to the
signs of terms $f_{d^2_i}$ with $i=1-5$, respectively.

\section{Role of four-fermion interaction in the low-energy states
for the symmetric situations}\label{Sec_interaction}

In the previous section~\ref{Sec_coupled_equations}, we have derived the
coupled flow equations for interaction parameters via clinching the
interplay among different four-fermion parameters and information of rotational and
particle-hole symmetries. In the spirt of RG, the low-energy behaviors of systems can
be conventionally extracted from these equations. Accordingly, we now are in a suitable
position to reveal the influence of interplay among distinct four-fermion couplings
on the low-energy properties of the 3D QBT systems.

It is well known that the order parameters or quasiparticles in the most of
condensed matter systems are inescapably coupled and mutually influenced due
to the strong quantum fluctuations nearby the QCP in the low-energy regime~\cite{Herbut2007Book,Sachdev1999Book,Fernandes2012PRB,Fernandes2013PRL}.
This immediately raises a question that whether and how the QCP is slightly
revised or completely changed by the interplay between distinct quartic couplings
in the low-energy regime?

\begin{figure}
\centering
\includegraphics[width=4.3in]{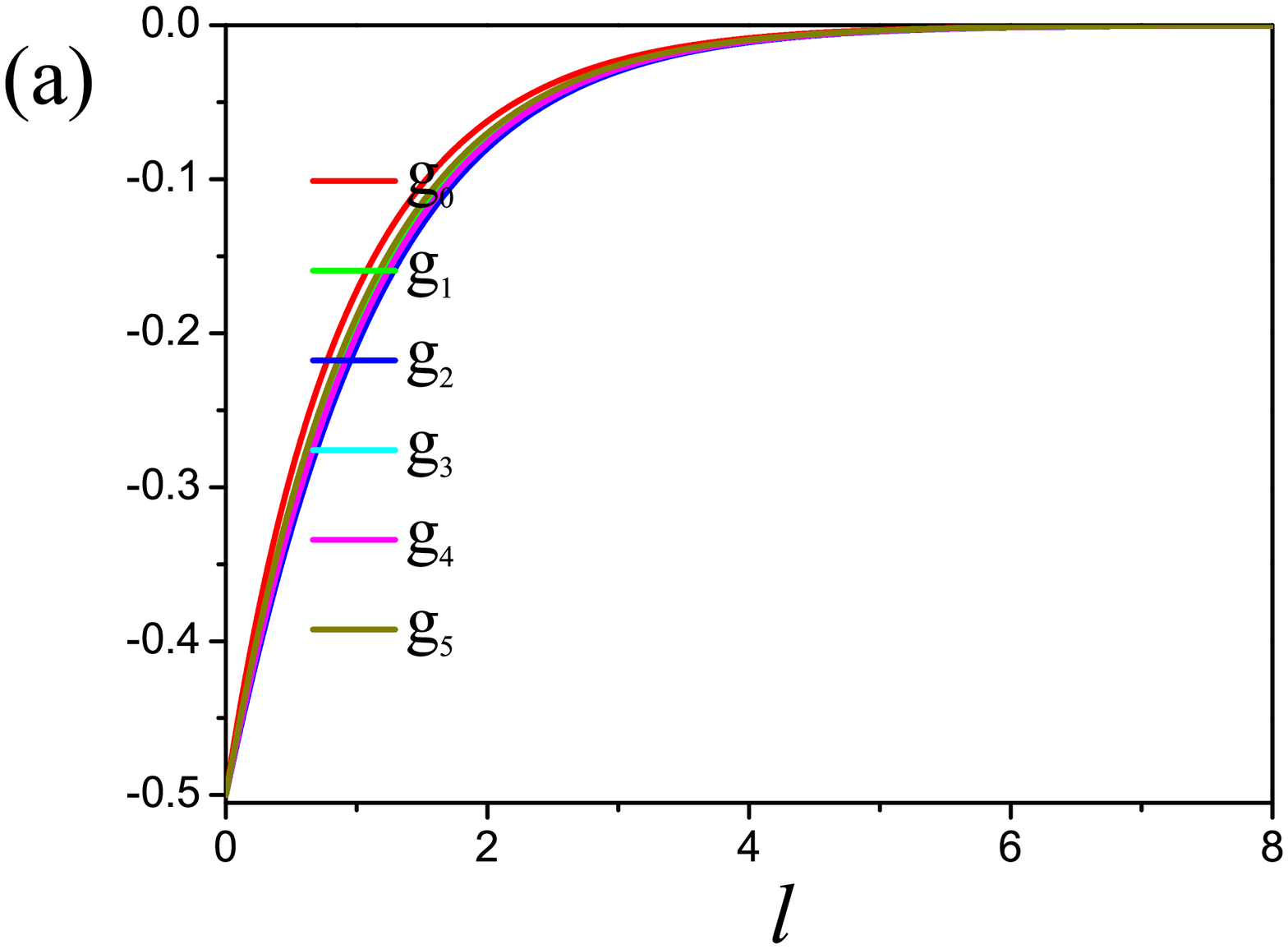}\vspace{-2.05cm}
\includegraphics[width=4.3in]{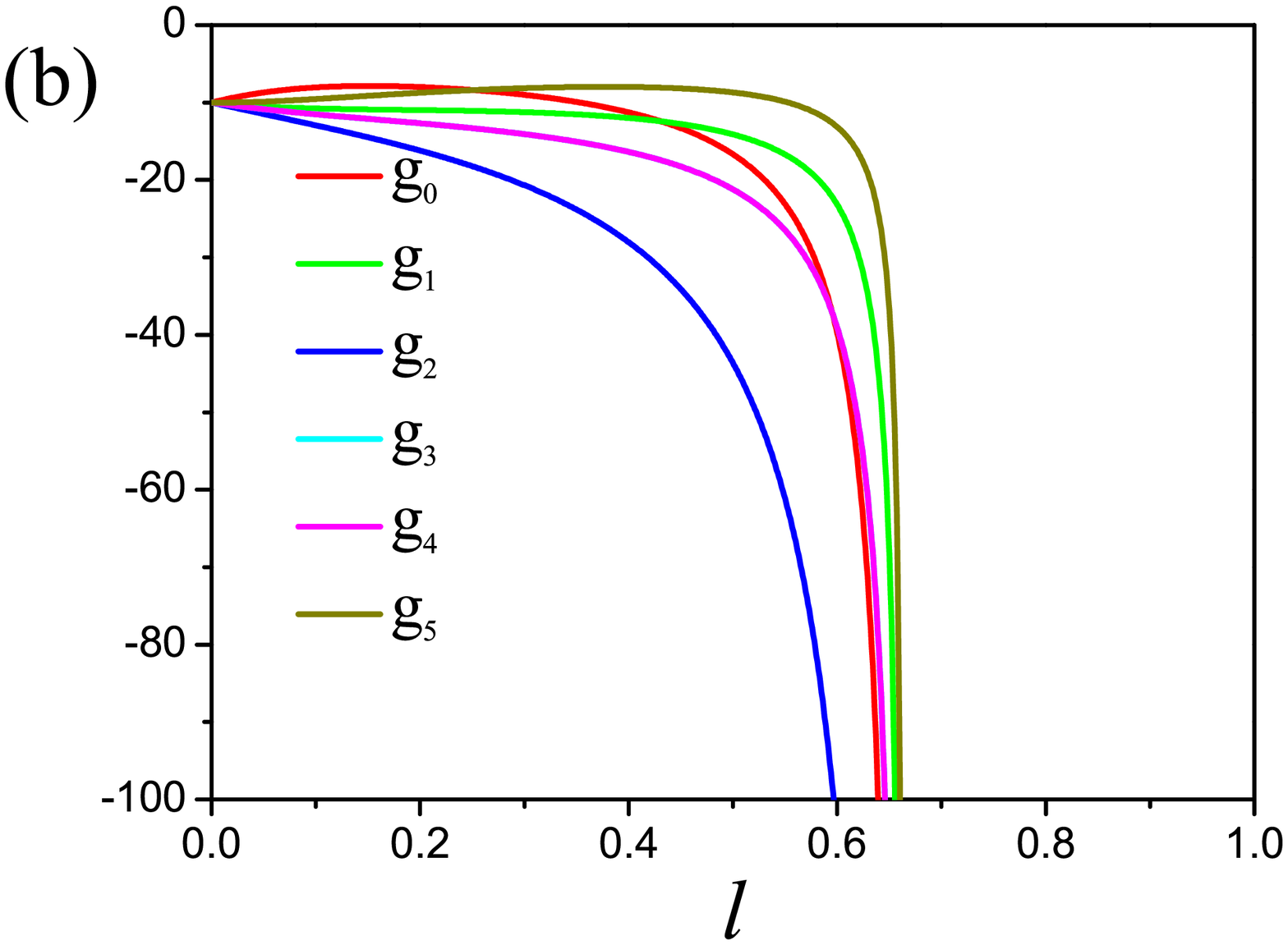}
\vspace{-2.1cm}
\caption{(Color online) Four-fermion couplings flow to: (a) the Gaussian fixed point for
small initial values with $g_i(l=0)<0$ (the basic results are the same for
$g_i(l=0)>0$ and not shown here); and (b) strong coupling for large
initial values with $g_i(l=0)<0$. All the energy-dependent evolutions
are measured by the $\Lambda^{-1}_0$. }\label{Fig_Symmetry}
\end{figure}

\begin{figure*}[htbp]
\centering
\includegraphics[width=3.05in]{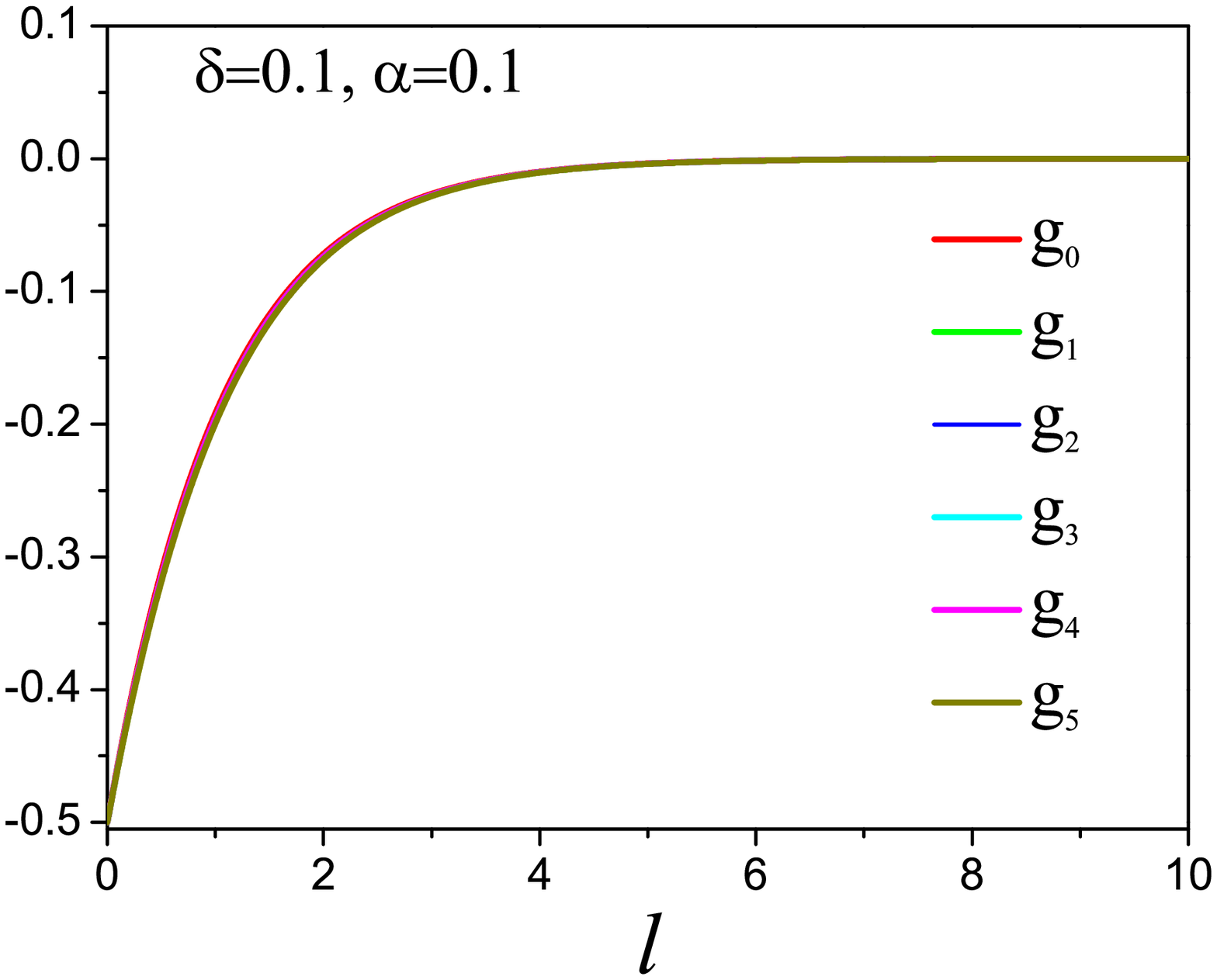}\hspace{-2.76cm}
\includegraphics[width=3.05in]{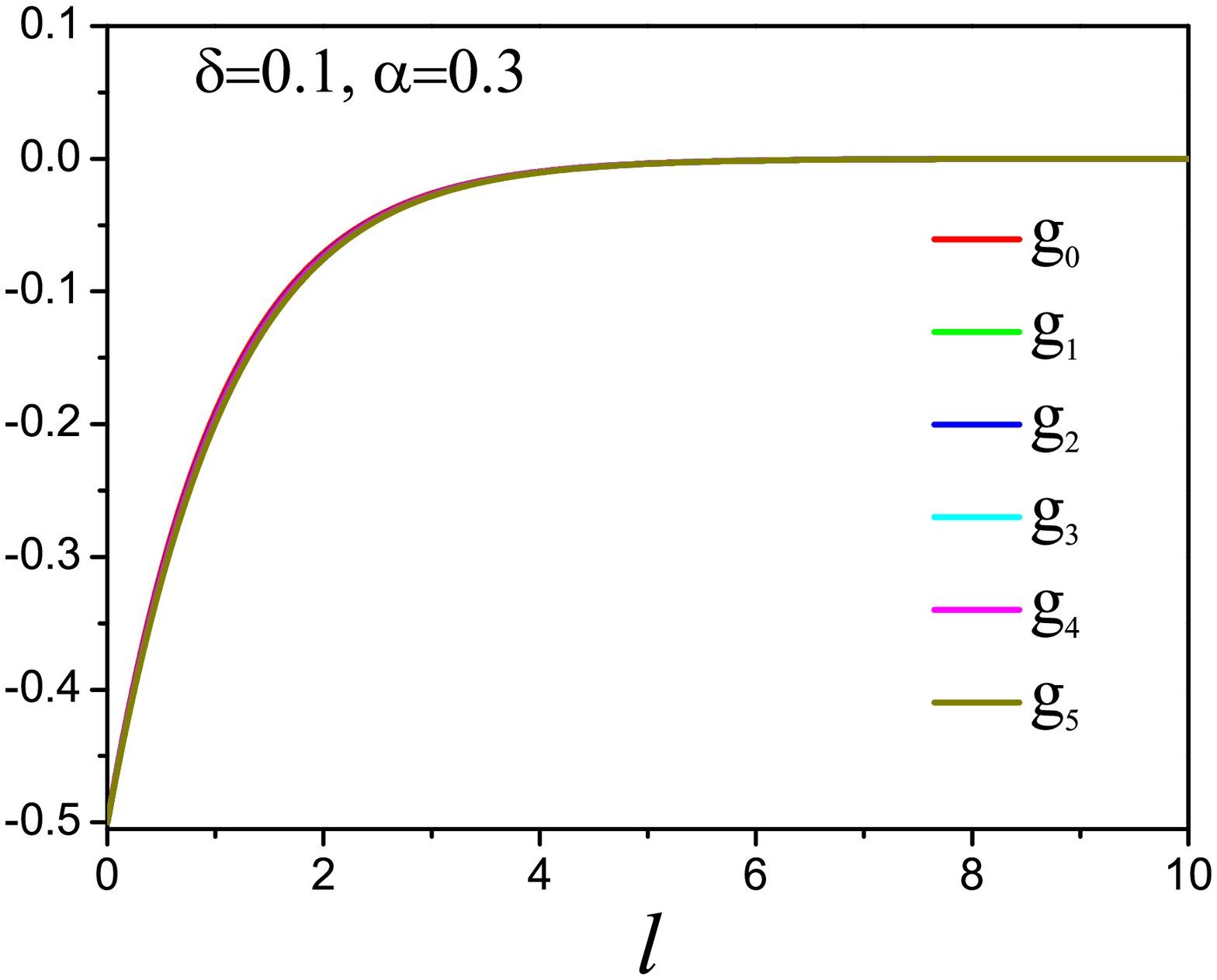}\hspace{-2.76cm}
\includegraphics[width=3.05in]{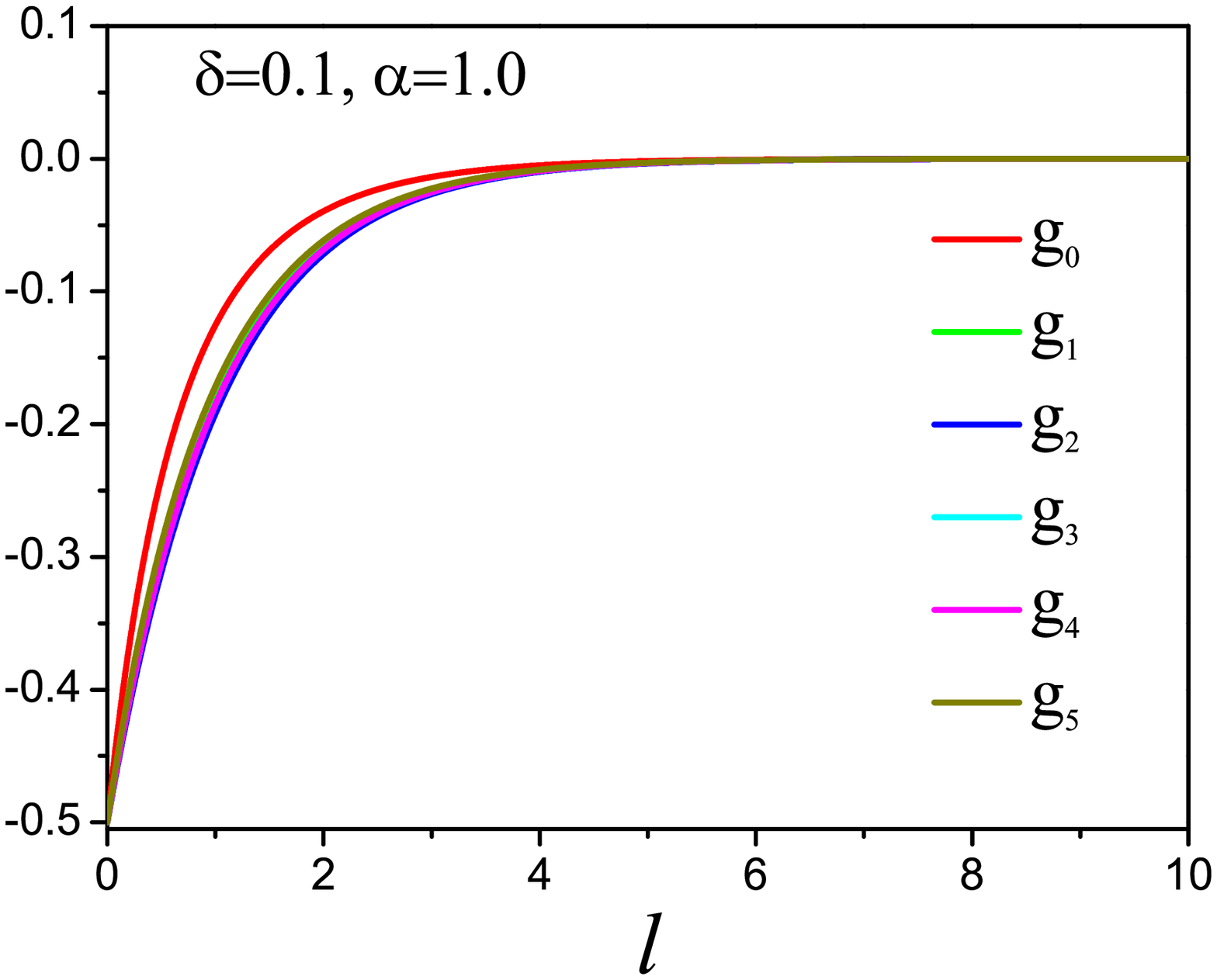}\\
\vspace{-1.5cm}
\includegraphics[width=3.05in]{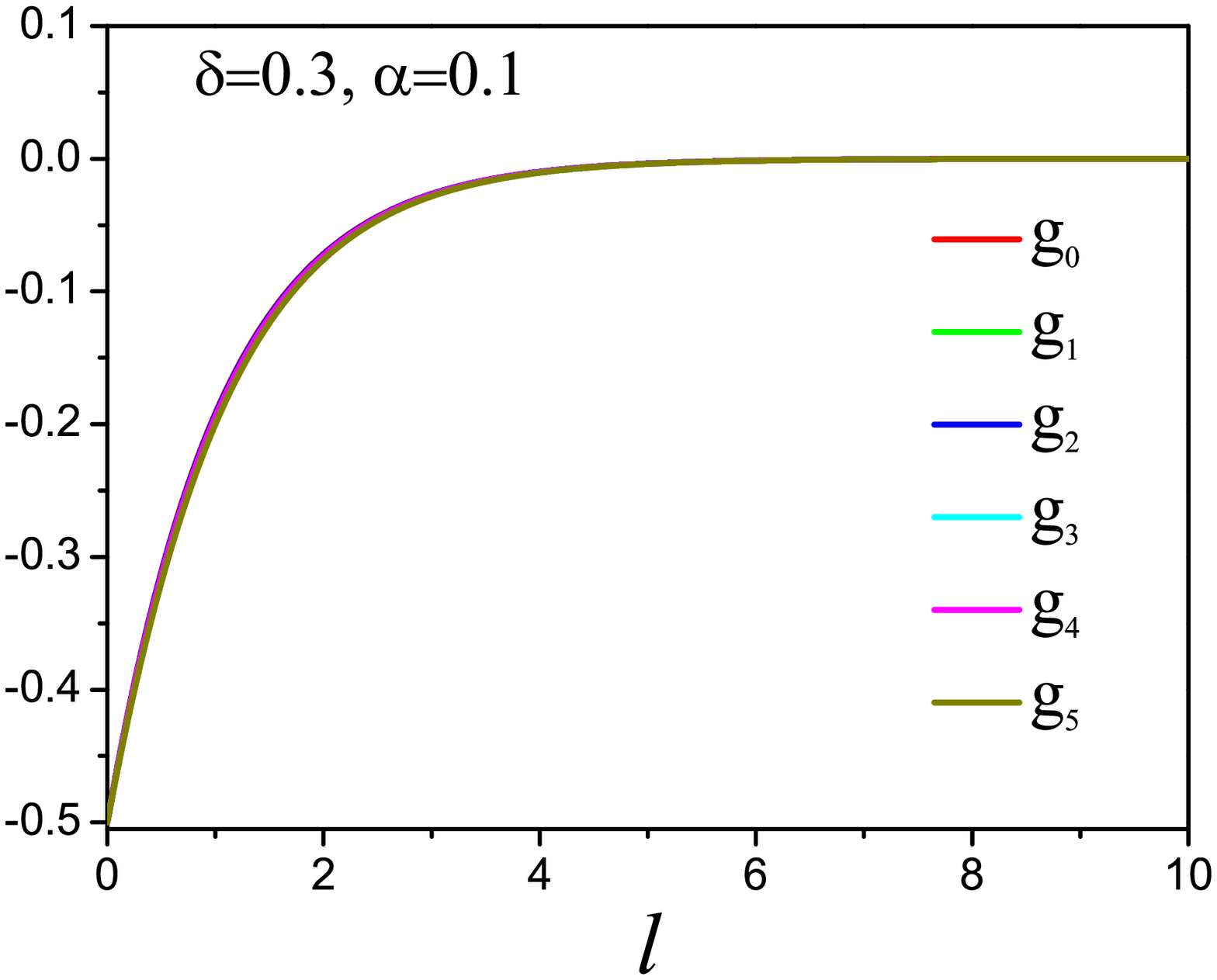}\hspace{-2.76cm}
\includegraphics[width=3.05in]{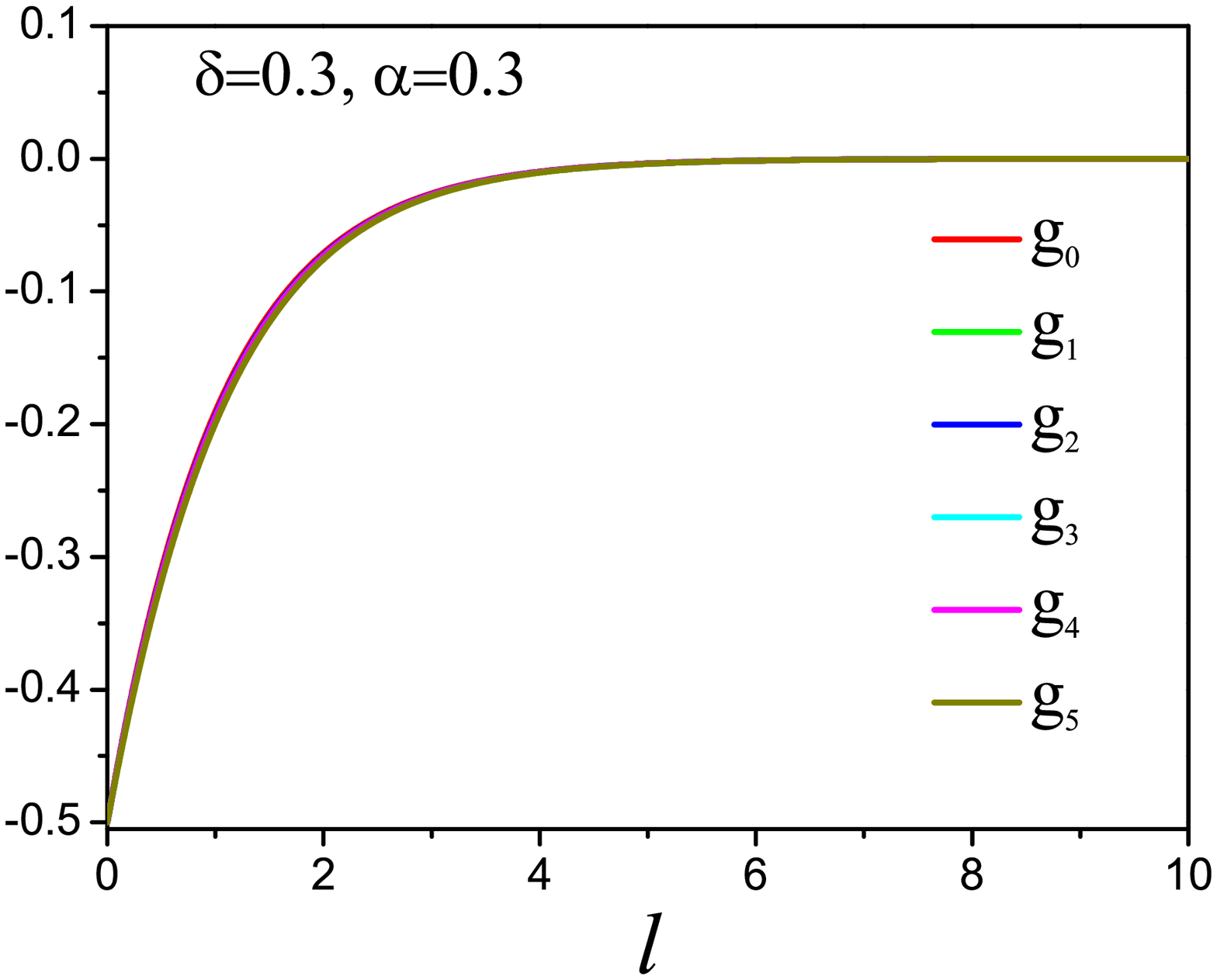}\hspace{-2.76cm}
\includegraphics[width=3.05in]{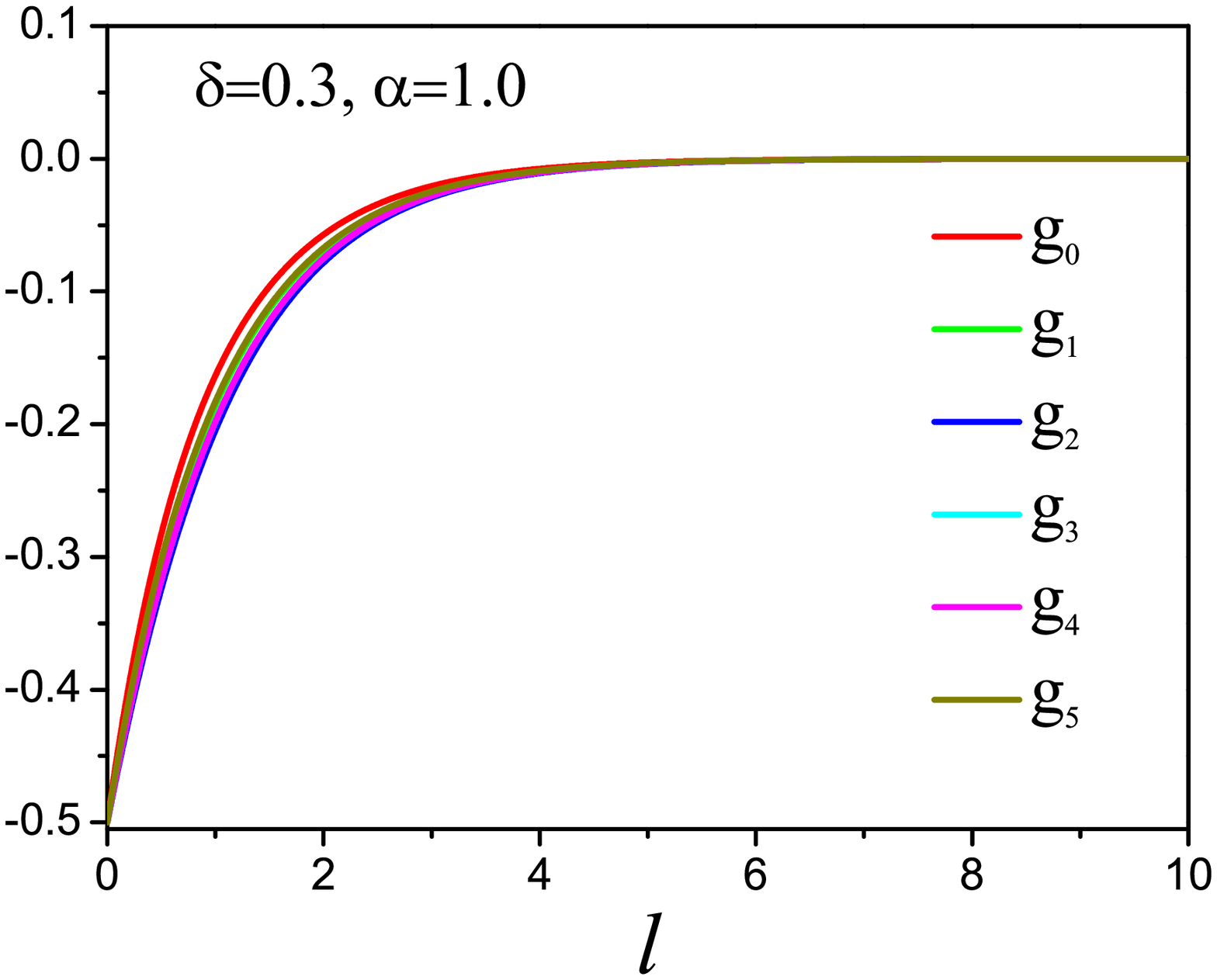}\\
\vspace{-1.5cm}
\includegraphics[width=3.05in]{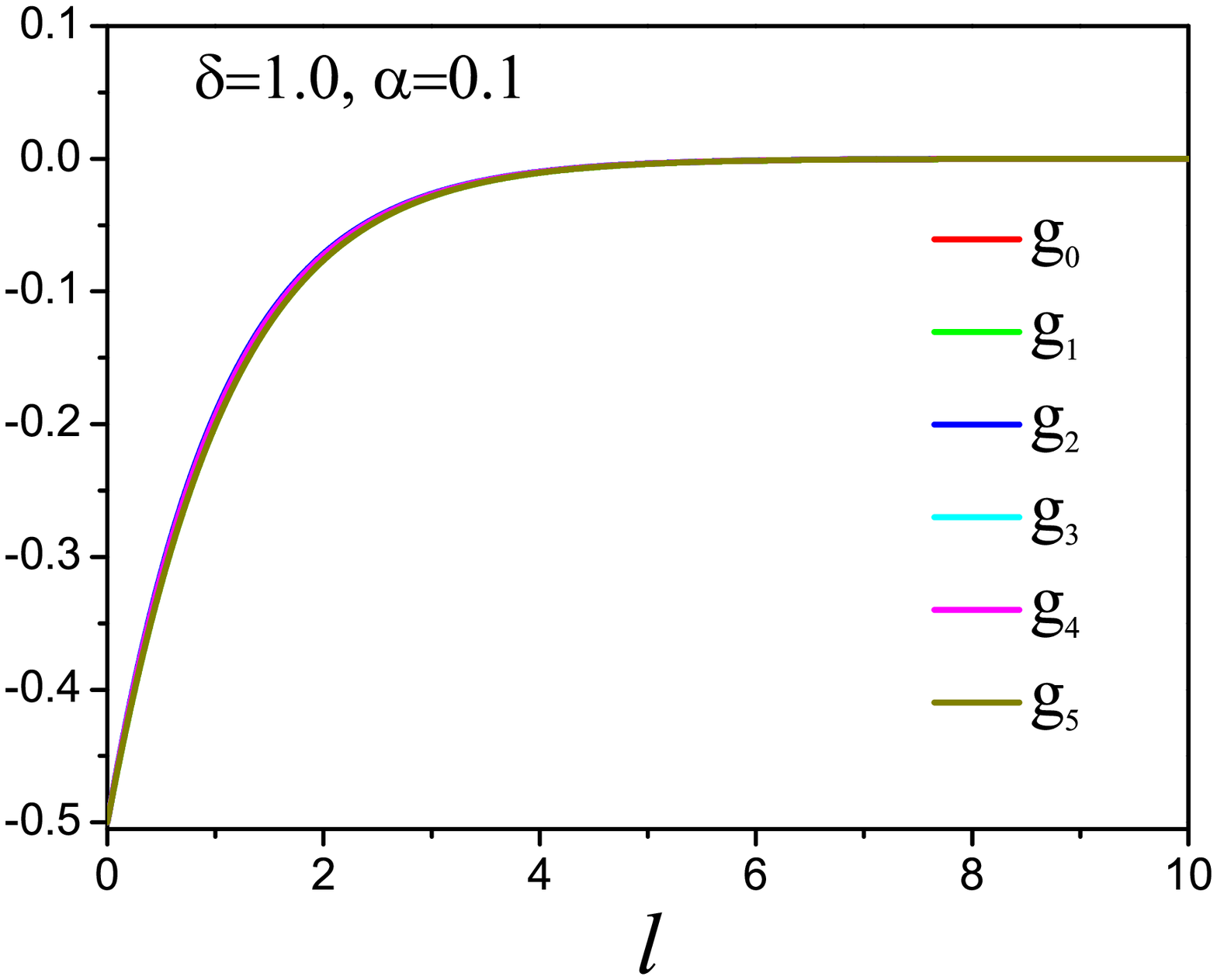}\hspace{-2.76cm}
\includegraphics[width=3.05in]{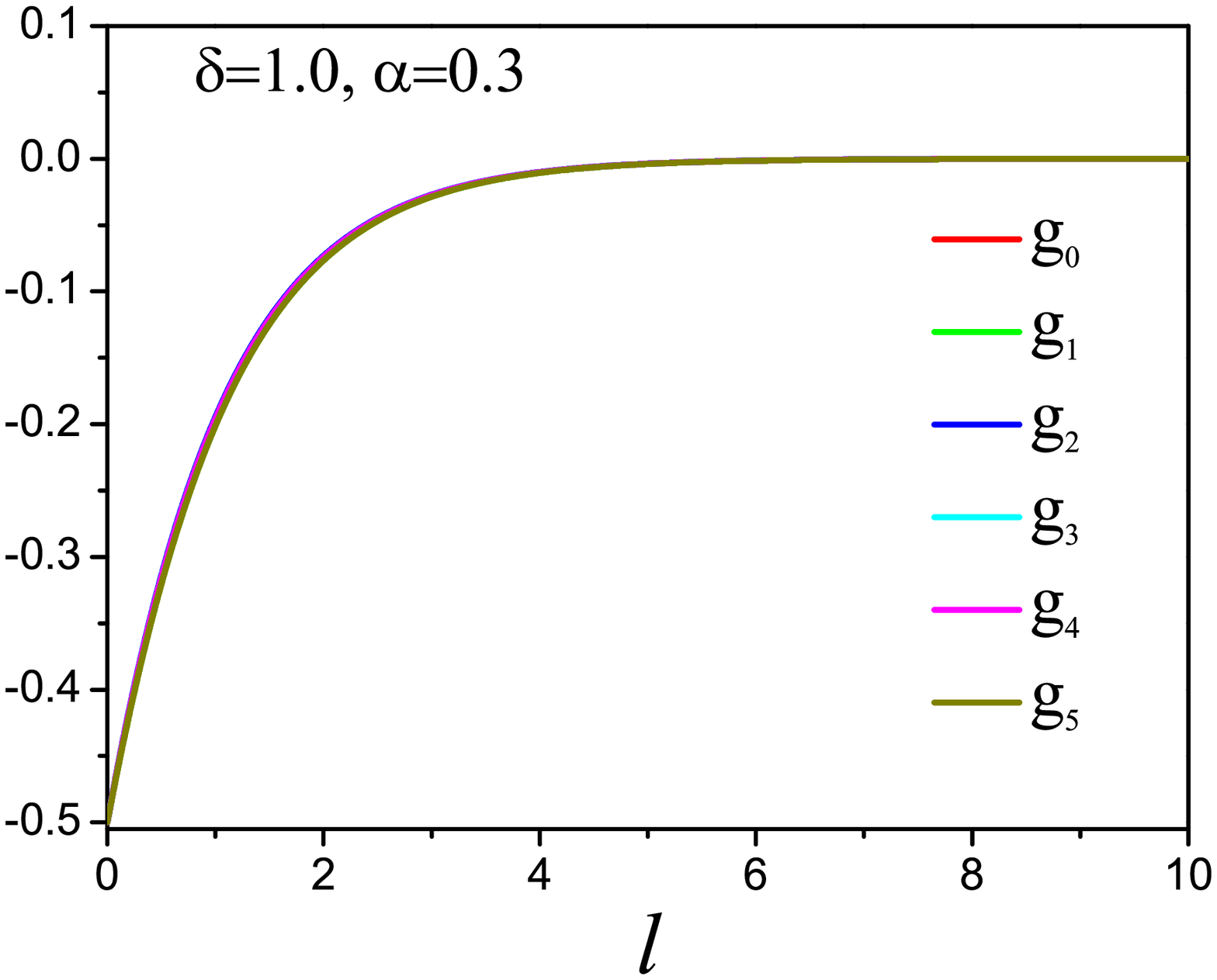}\hspace{-2.76cm}
\includegraphics[width=3.05in]{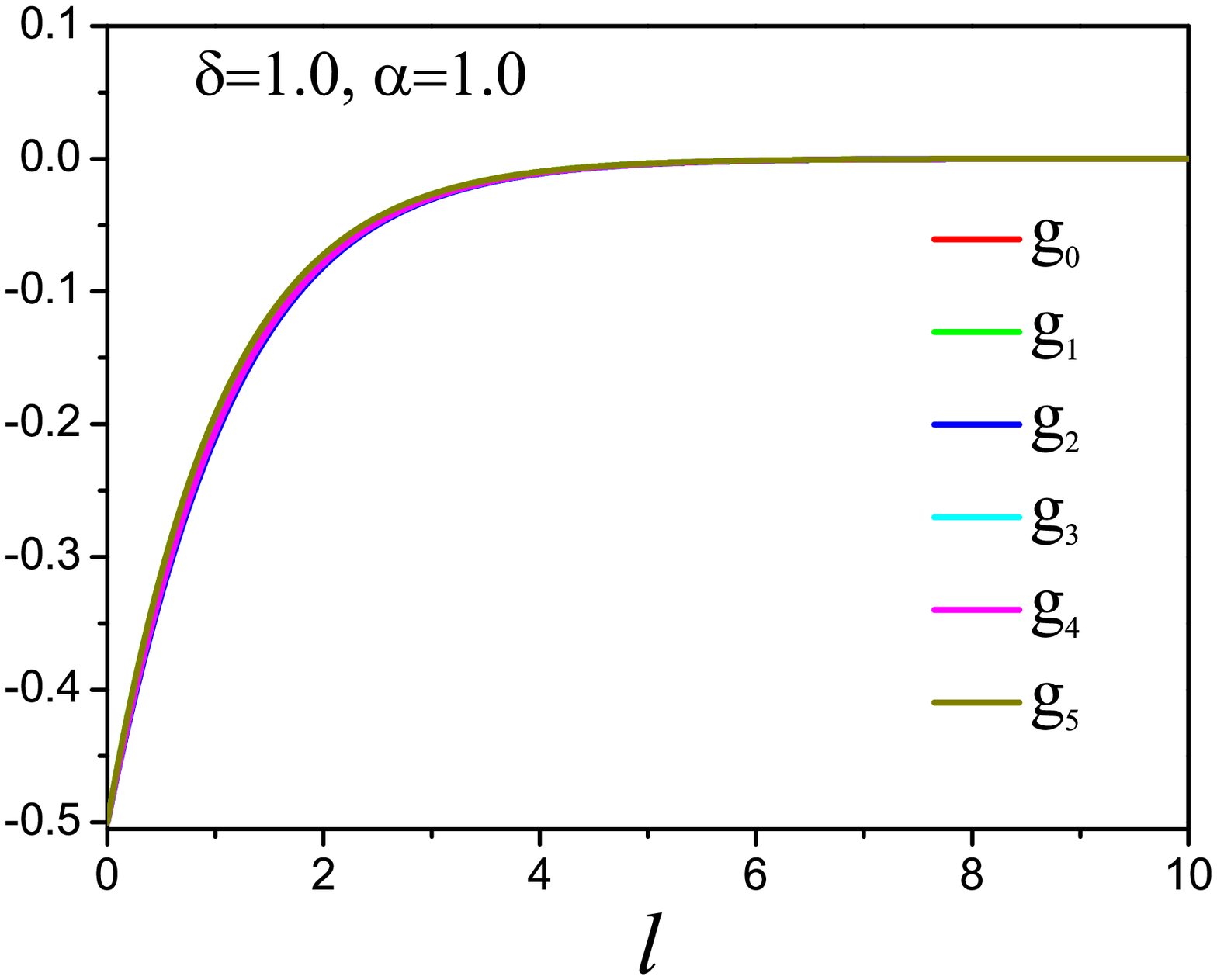}\\
\vspace{-1.2cm}
\caption{(Color online) Evolutions of four-fermion couplings for the
small initial values with $g_i(l=0)<0$ with $i=0-5$ (the basic results are the same for
$g_i(l=0)<0$) and several representative values of asymmetric parameters
$\delta$ and $\alpha$. All the energy-dependent evolutions
are measured by the $\Lambda^{-1}_0$ (the flows of couplings $g_2$ and $g_3$
are nearly overlapped).}\label{Fig_delta_xi_small_minus}
\end{figure*}

\begin{figure*}[htbp]
\centering
\includegraphics[width=3.05in]{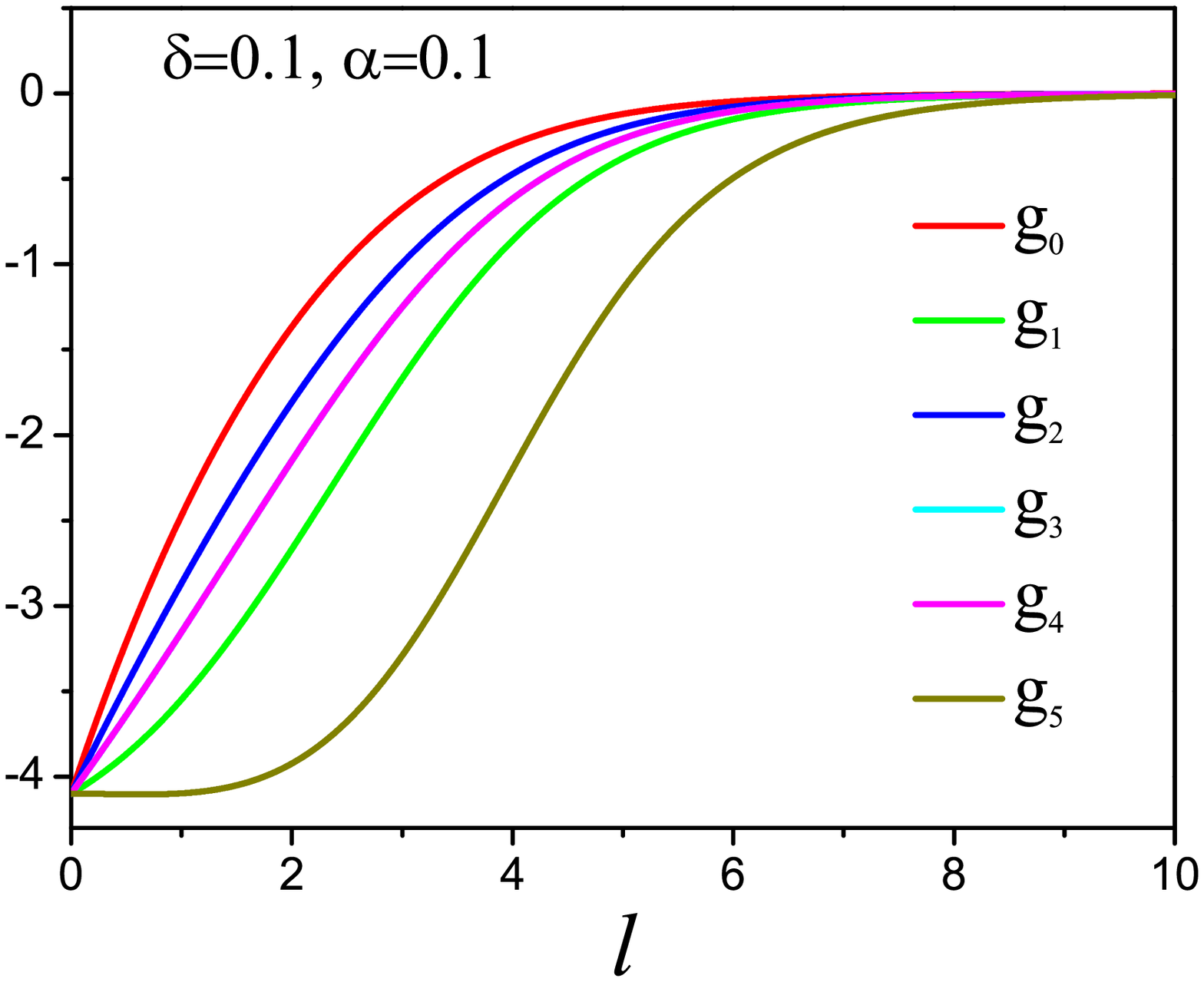}\hspace{-2.76cm}
\includegraphics[width=3.05in]{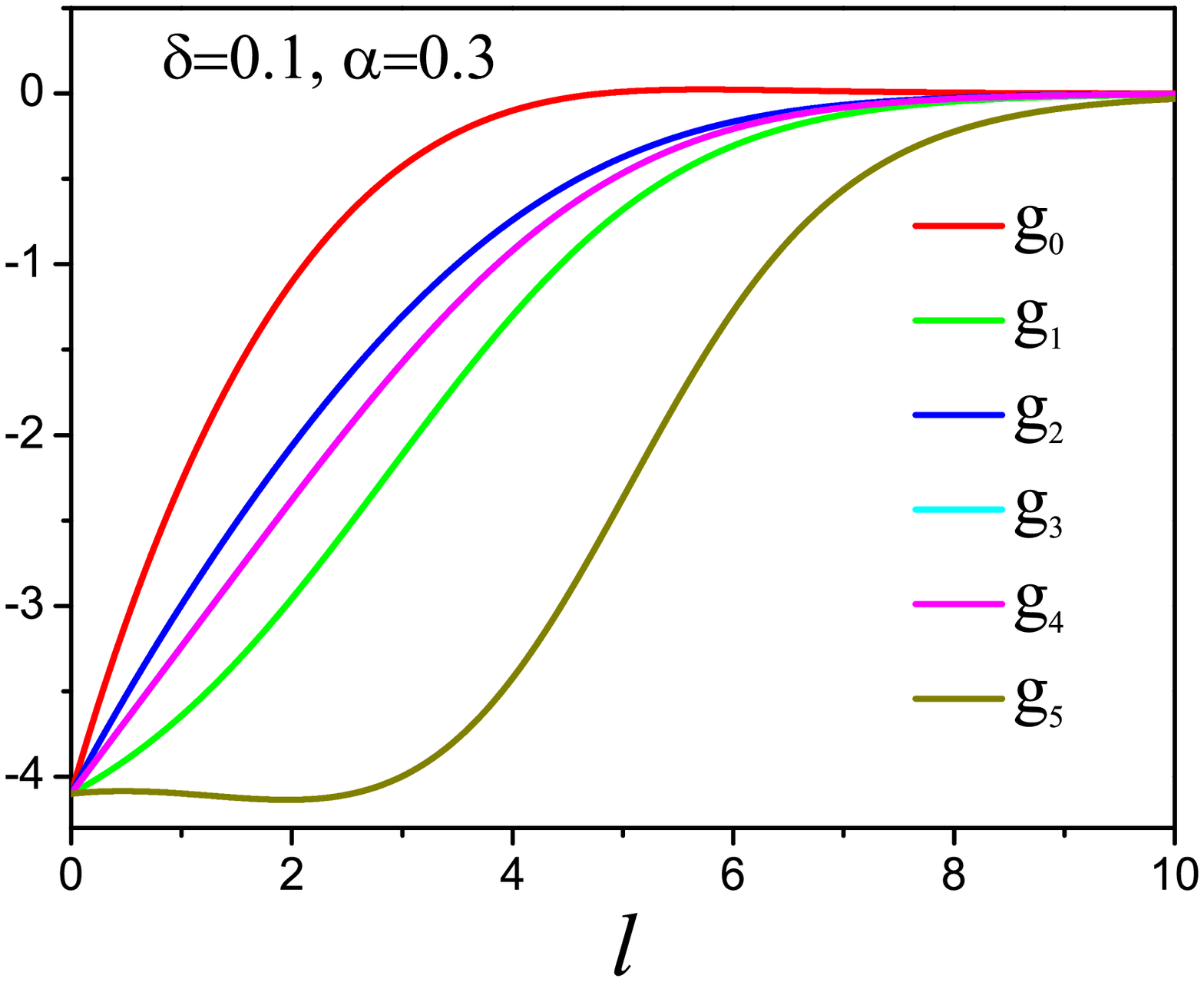}\hspace{-2.76cm}
\includegraphics[width=3.05in]{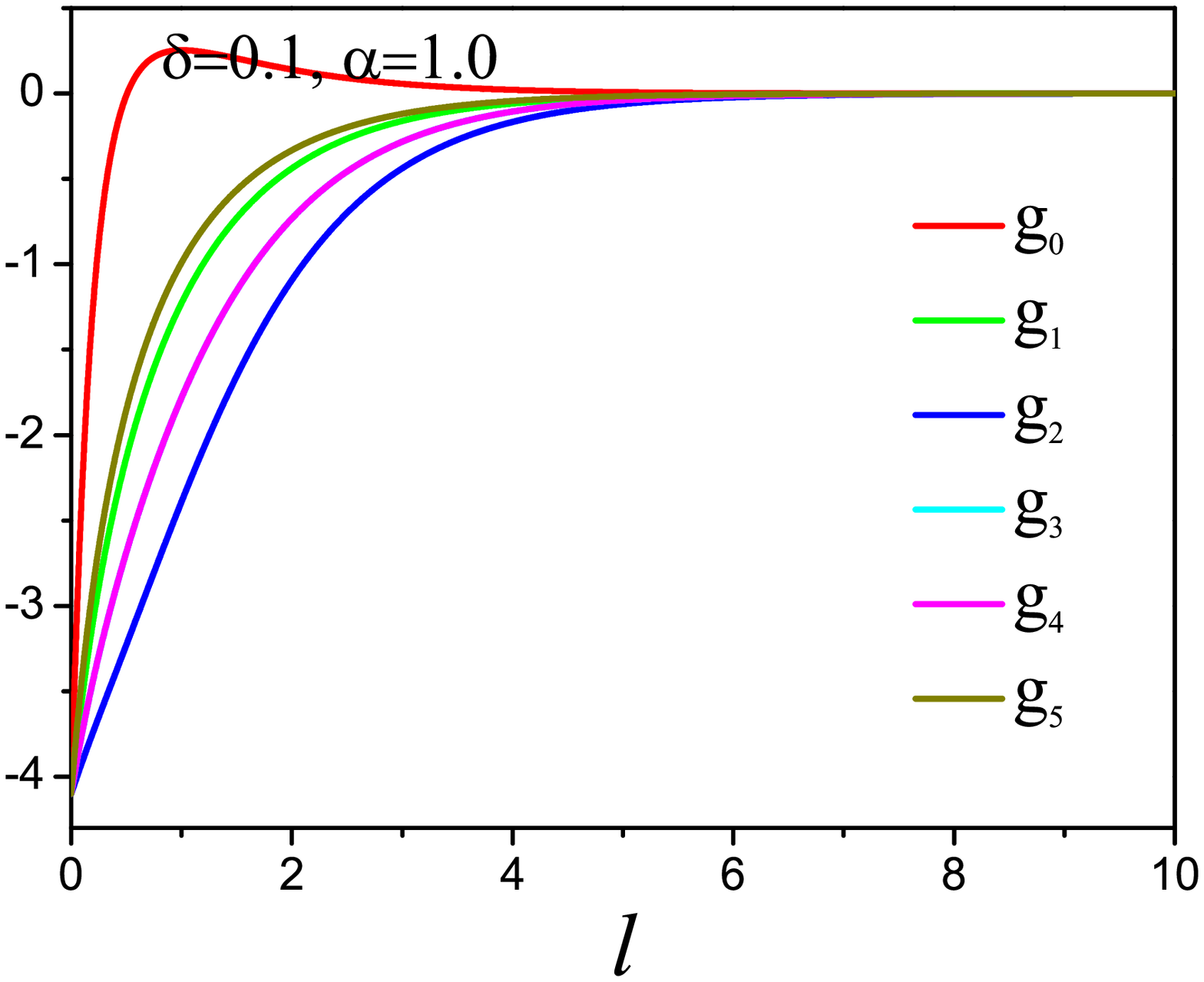}\\
\vspace{-1.5cm}
\includegraphics[width=3.05in]{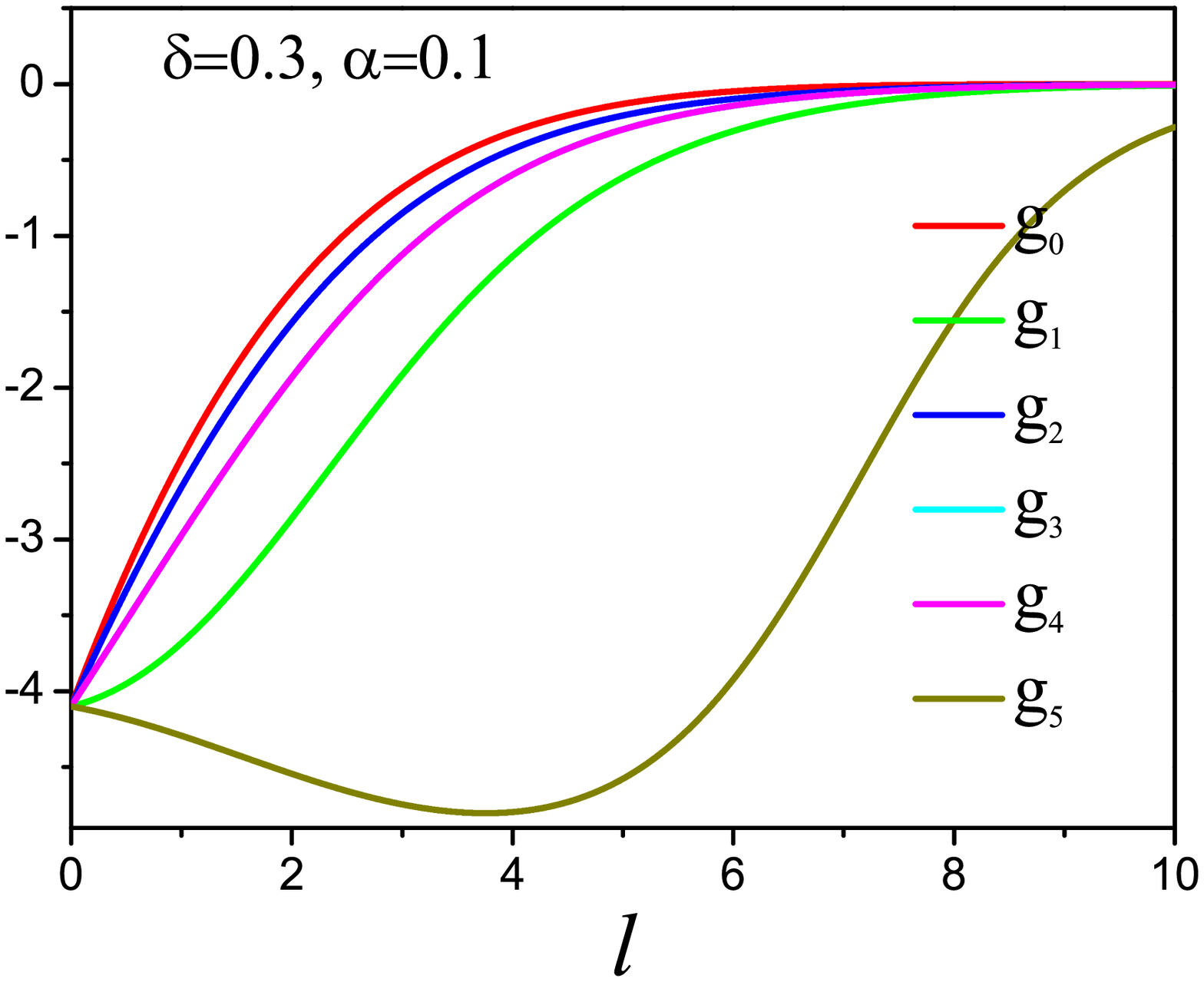}\hspace{-2.76cm}
\includegraphics[width=3.05in]{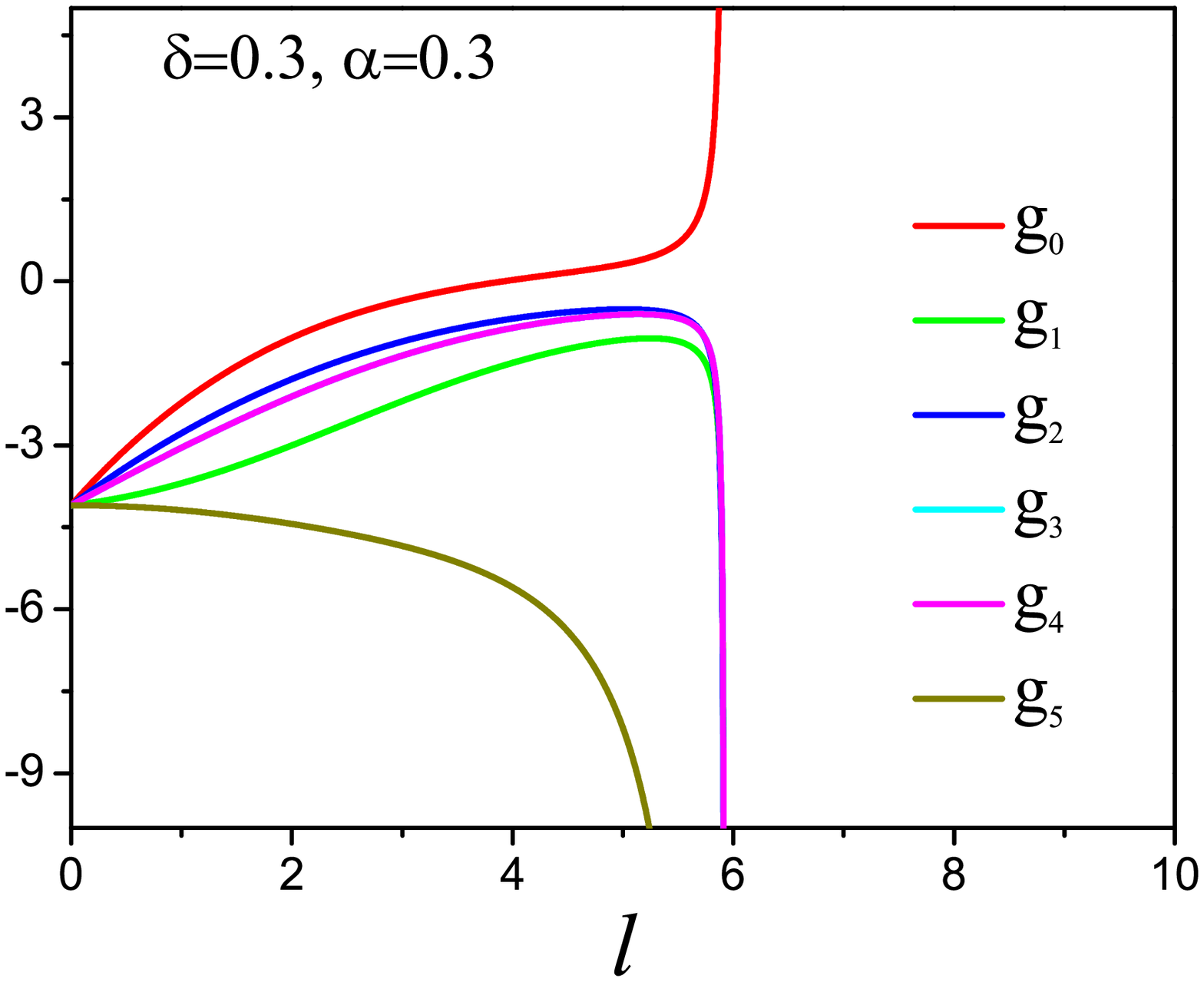}\hspace{-2.76cm}
\includegraphics[width=3.05in]{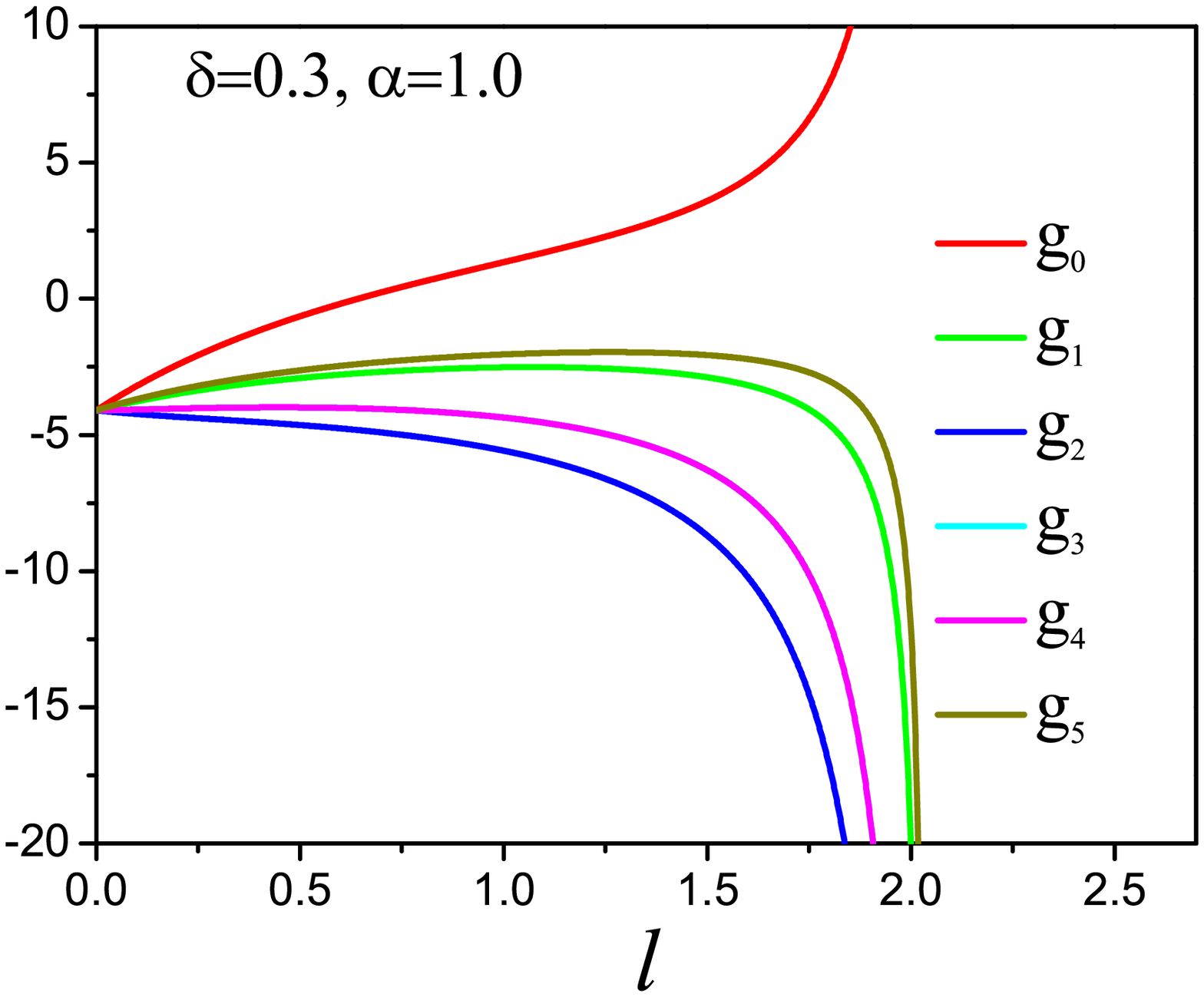}\\
\vspace{-1.5cm}
\includegraphics[width=3.05in]{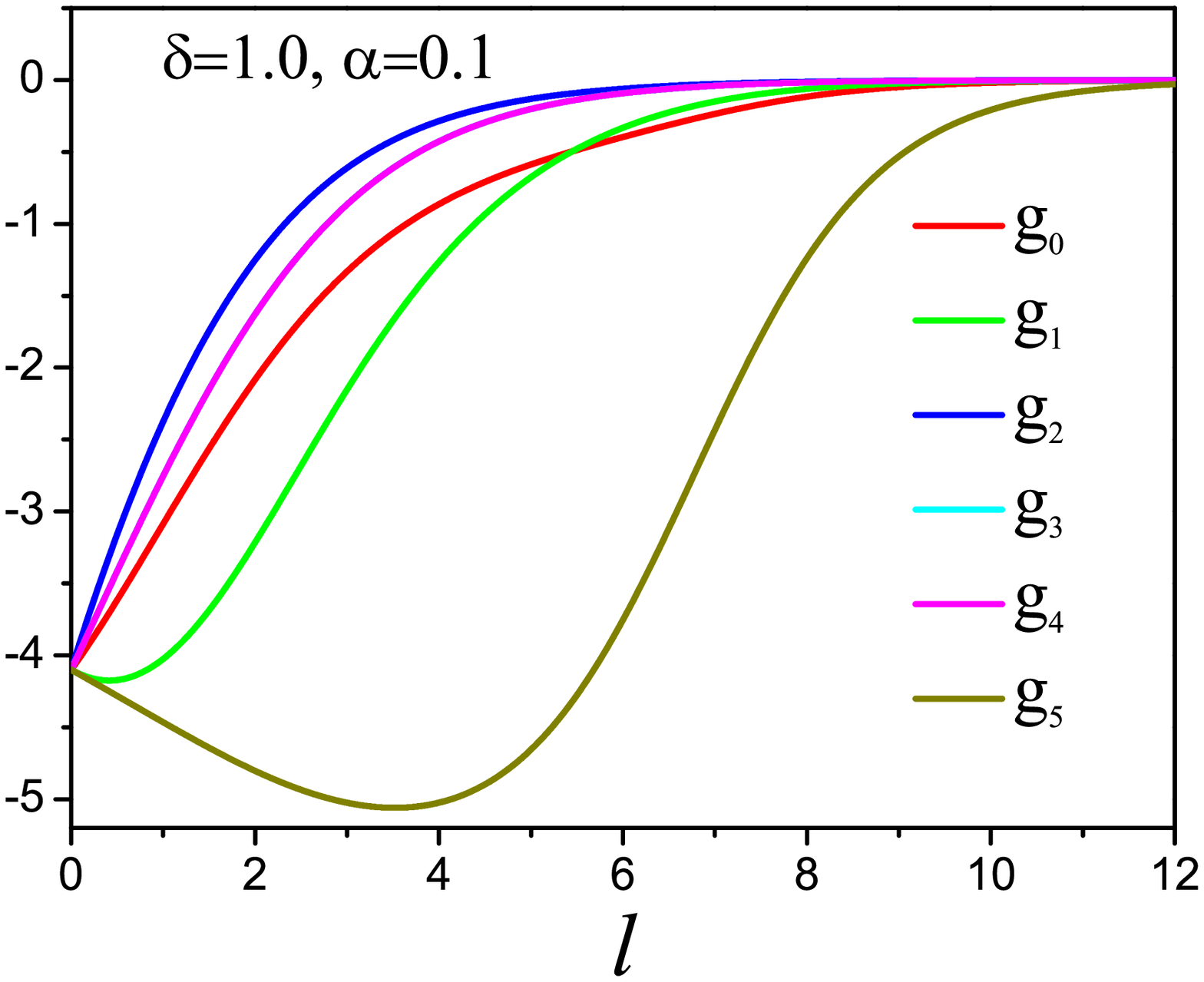}\hspace{-2.76cm}
\includegraphics[width=3.05in]{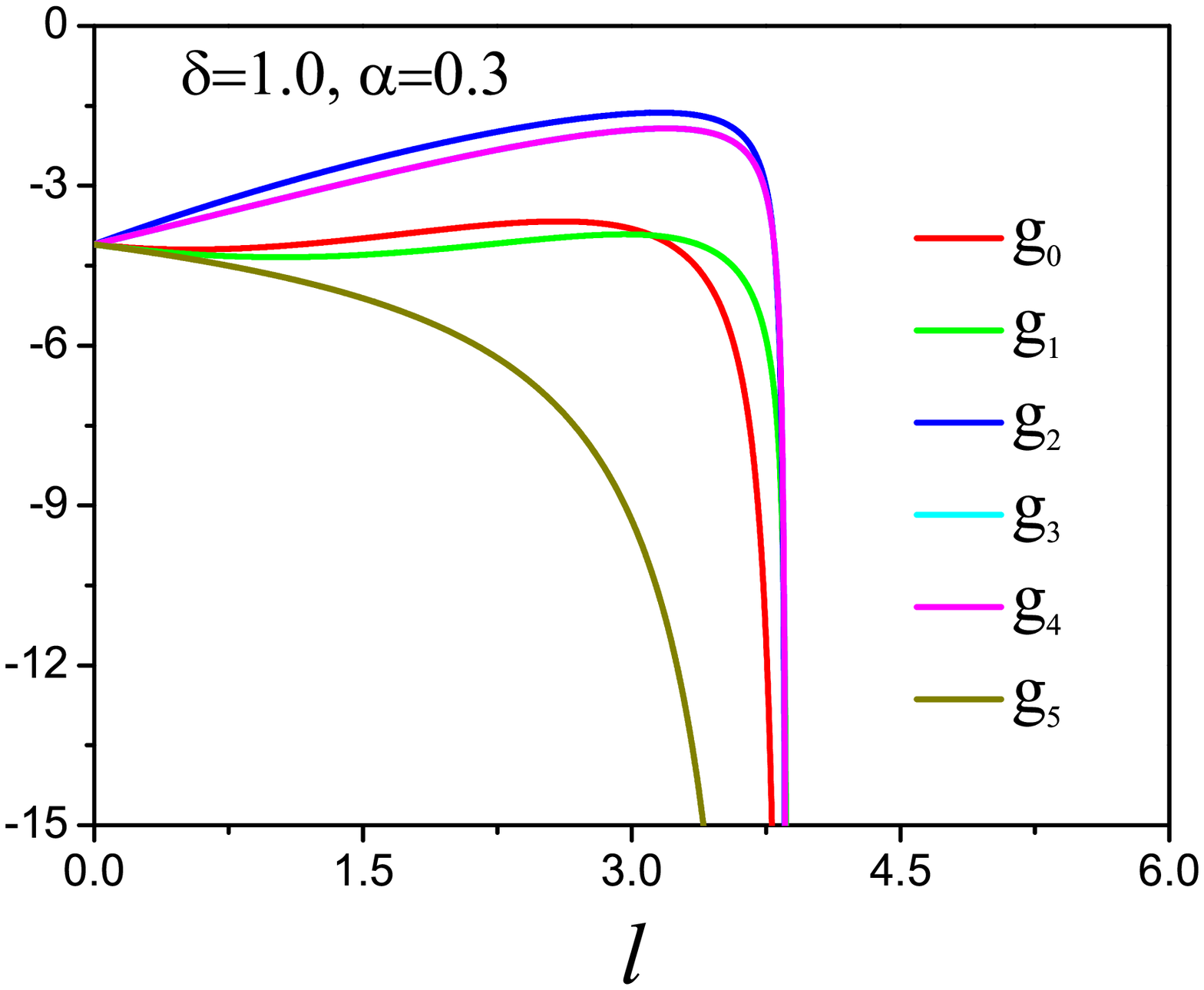}\hspace{-2.76cm}
\includegraphics[width=3.05in]{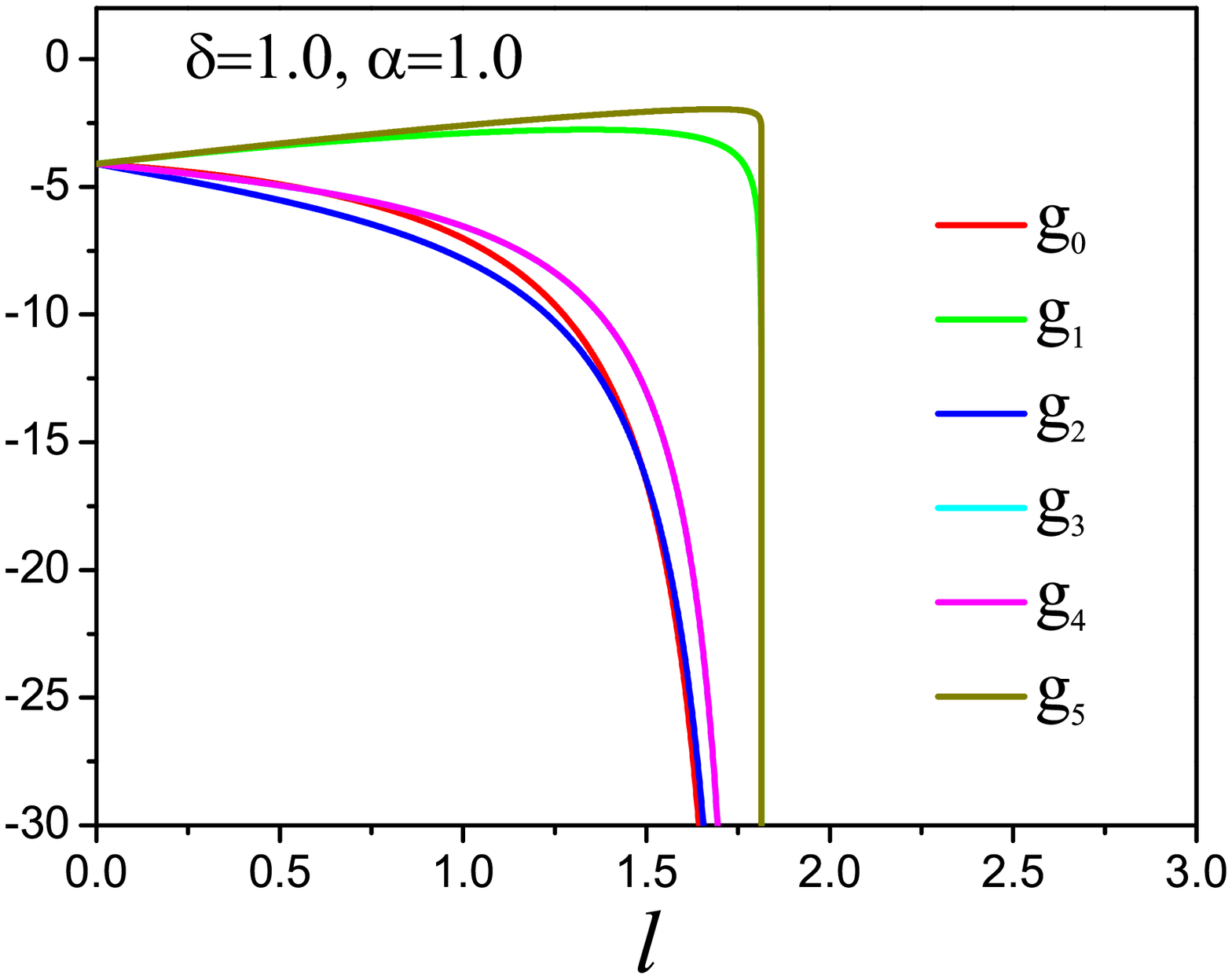}\\
\vspace{-1.2cm}
\caption{(Color online) Evolutions of four-fermion couplings for the
moderate initial values with $g_i(l=0)<0$ and distinct values of asymmetric parameters
$\delta$ and $\alpha$. All the energy-dependent evolutions
are measured by the $\Lambda^{-1}_0$ (the flows of couplings $g_2$ and $g_3$
are nearly overlapped).}\label{Fig_delta_xi_large_minus}
\end{figure*}

Before going further, we would like stress that there exists two
distinct situations, which are distinguished by presence or absence of the rotational and particle-hole
symmetries. For the asymmetric cases, the rotational and/or particle-hole symmetries would be broken and the
asymmetries are representatively measured by the parameters $\alpha$ and $\delta$, with
which the associated flow equations of four-fermion couplings are presented
in Eq.~(\ref{Eq_flow_Eq_g_0})-(\ref{Eq_flow_Eq_g_5}). In a sharp contrast,
the system preserves both the rotational and particle-hole symmetries for
the symmetric situations, which we put our focus on within this section.
To proceed, the corresponding evolutions are given by Eqs.~(\ref{Eq_flow_Eq_g_0_xi_0})-(\ref{Eq_flow_Eq_g_5_xi_0}). These fermion-fermion parameters in Eq.~(\ref{Eq_model}), $g_i$ with $i=0$ to $5$,
manifestly become energy-dependent and are restricted by each other upon lowering the
energy scale.  After incorporating into the intimate influence of these fermionic couplings,
the correlatedly four-fermion parameters in Eq.~(\ref{Eq_model}), $g_i$, $i=0-5$,
become closely energy-dependent and are restricted by coupled flow equations upon lowering the
energy scale. With the respect to the rotational and particle-hole
symmetries, we are informed that the parameters $\alpha=\alpha_s$ and $\delta=\delta_s$. Concretely, we
utilize the corresponding RG equations ~(\ref{Eq_flow_Eq_g_0_xi_0})-(\ref{Eq_flow_Eq_g_5_xi_0})
and arrive at the results as depicted in Fig.~\ref{Fig_Symmetry} after performing
the straightforwardly numerical calculations. Studying from Fig.~\ref{Fig_Symmetry}(a),
we obtain that all four-fermion couplings flow towards zero
at the lowest energy limit if the initial values of $|g_i(l=0)|$, $i=0-5$, are
sufficiently small. This clearly indicates that the system under some starting
situations runs to the Gaussian fixed point at the lowest energy limit although the
mutual influence among distinct coupling are switched on. However, the Gaussian fixed
point is no longer stable and qualitatively moved by the four-fermion interactions
if the initial values of $|g_i(l=0)|$, $i=0-5$, are adequately large as
demonstrated in Fig.~\ref{Fig_Symmetry}(b). To be specific, all the couplings
parameters flows away from the Gaussian fixed point. In particular, it is interesting to
point out the evolutions of $g_i$ with $i=1-5$ are overlapped due to their RG equations
taking the similar structure as shown in Eq.~\ref{Eq_flow_Eq_g_5_xi_0}. Gathering the
information of this subfigure,
we unambiguously find that the Gaussian fixed point is changed qualitatively and
all the four-fermion parameters $g_i$, $i=0-5$, flow towards the strong couplings at the
lowest-energy limit.

To be brief, it is worth pointing out that there are several interesting
points are triggered by the four-fermion interactions.
In variance with the noninteracting circumstance, the Gaussian fixed point can either
be reached if the initial values of fermionic parameters are small or
qualitatively destroyed and flow towards strong coupling when their starting values
are sufficiently large. Therefore, the QCP in noninteracting case is erased and
replaced by the strong couplings if the four-fermion interactions are turned on.

\section{Effects of interactions combined with asymmetries on the low-energy states}\label{Sec_interaction_asymm}

As presented in previous section, the low-energy states of 3D QBT somehow, even for
the preservation of rotational and particle-hole symmetries, can be revised by the
mutual interactions among different types of four-fermion couplings. One maybe naturally
ask further what about the roles of rotational and particle-hole asymmetries in the low-energy
behaviors meanwhile these four-fermion interactions are switched on. We are going to investigate
these in the rest of this section.

\subsection{Split evolutions of interaction parameters}

In order to study the general case in which the rotational and particle-hole
symmetries cannot always be preserved, we are forced to transfer our focus from the coupled flow equations Eqs.~(\ref{Eq_flow_Eq_g_0_xi_0})-(\ref{Eq_flow_Eq_g_5_xi_0}) to the asymmetric situations,
namely Eqs.~(\ref{Eq_flow_Eq_g_0})-(\ref{Eq_flow_Eq_g_5}), which are derived
from the presence of general values of $\alpha$ and $\delta$. By carrying out the
analogous procedures, we respectively summarized the primary results in Fig.~\ref{Fig_delta_xi_small_minus}
and Fig.~\ref{Fig_delta_xi_large_minus} for small and large initial values of
four-fermion couplings and several selected values of asymmetric parameters.

We here would like to discuss the behaviors of trajectories for fermionic couplings
influenced by $\alpha$ and $\delta$, leaving the study of fixed points in the
following subsection. Collecting the information of Fig.~\ref{Fig_delta_xi_small_minus}
and Fig.~\ref{Fig_delta_xi_large_minus}, we find that, although the final destination may
be not changed, the energy-dependent trajectories of fermionc couplings are stretched to be
split no matter their starting values are small or large. These behaviors are sharply compared
to the overlapped trajectories in noninteracting case, in particular for small starting values
of four-fermion couplings as compared to Fig.~\ref{Fig_Symmetry}(a).
This is apparently delineated in Fig.~\ref{Fig_delta_xi_small_minus} and the separations
among these five distinct interaction parameters are evidently broadened upon increasing the
values of $\alpha$ and $\delta$. Moreover, the split distance is increasingly separated
as the asymmetric parameters $\alpha$ and $\delta$ are tuned to be large.
Based on these, we therefore ascribe this split of fermion-fermion
couplings to the role of rotational and particle-hole asymmetries.

\begin{figure}
\centering
\includegraphics[width=4.39in]{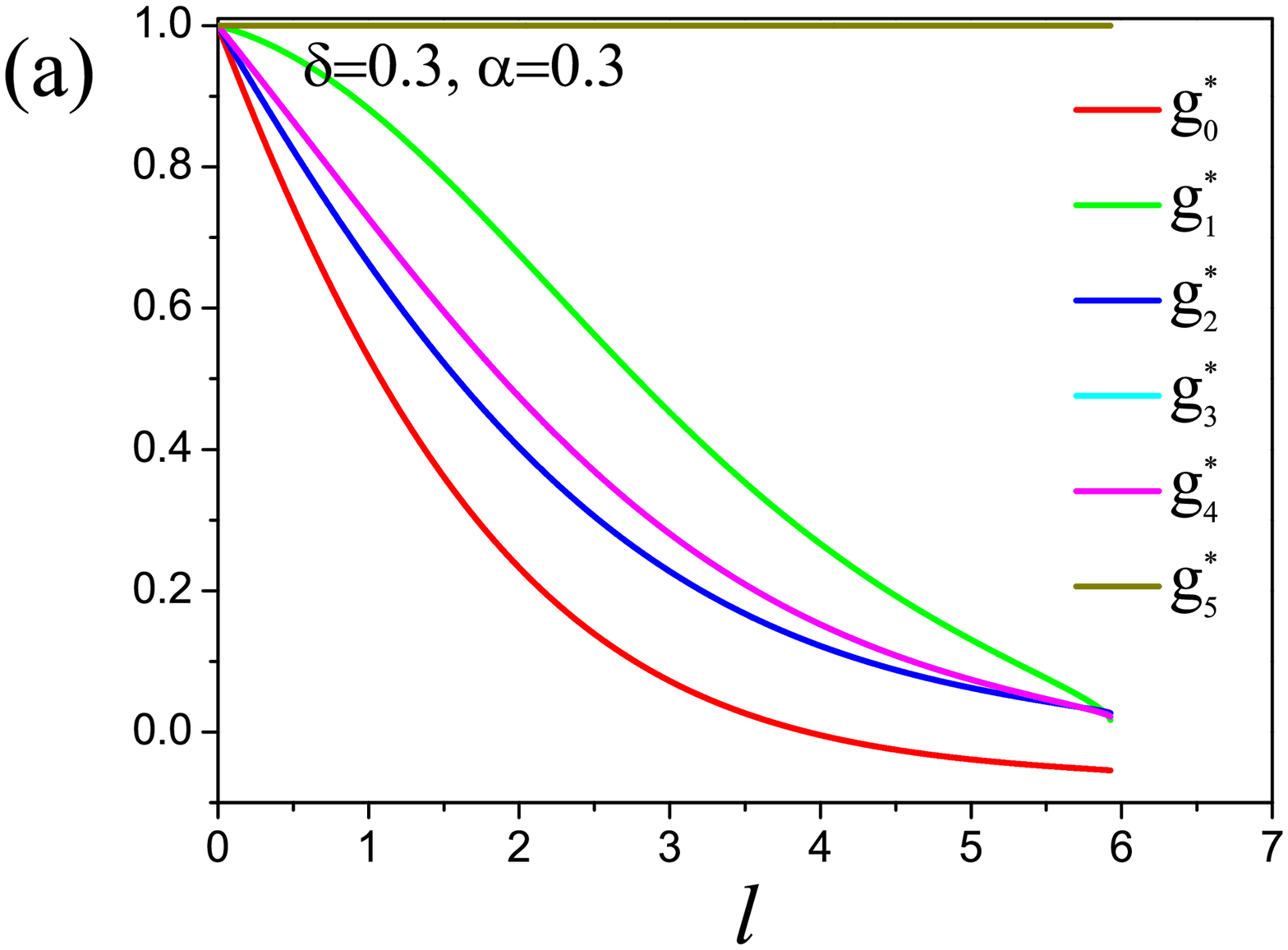}\vspace{-2.0cm}
\includegraphics[width=4.39in]{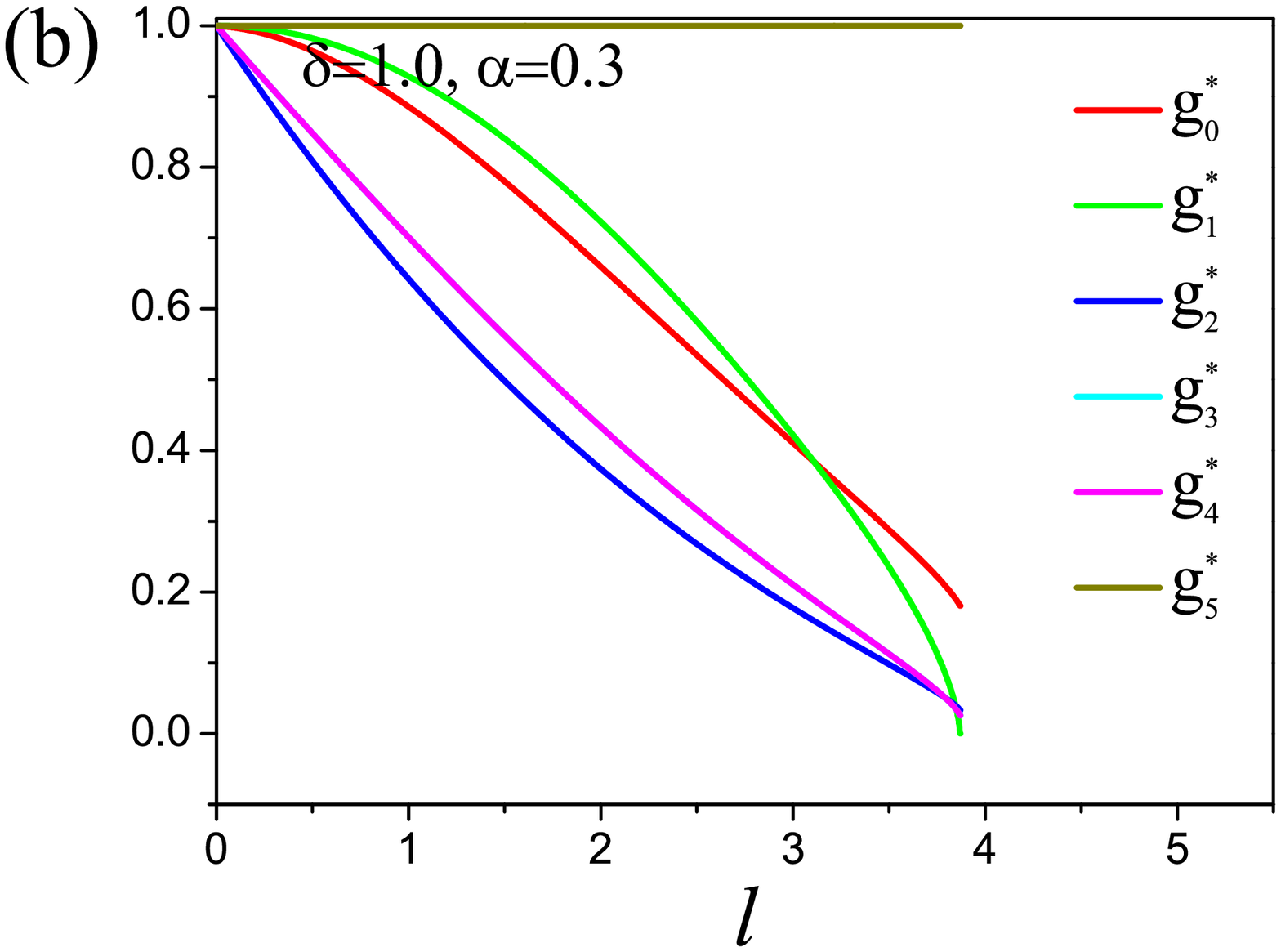}
\vspace{-2.0cm}
\caption{(Color online) Relatively fixed point at the lowest-energy limit at some
representative values of asymmetric parameters: (a) $\delta=0.3$ and $\alpha=0.3$
and (b) $\delta=0.3$ and $\alpha=1$. The rescaled
parameters $g^*_i(l)=g_i(l)/g_5(l)$, $i=0,1,2,3,4,5$ (the flows of couplings $g_3$ and $g_4$
are nearly degenerated).}\label{Fig_rescaled_FPs}
\end{figure}

\subsection{Instability}

As already presented in Sec.~\ref{Sec_interaction}, the rotational and particle-hole
asymmetries can play an important role in the low-energy behaviors. However, it is necessary
to collect the contribution from both the four-fermion interactions and asymmetries, which
together determine the fate of 3D QBT system at the lowest-energy limit.

At the start, we assume the initial values of fermionic couplings are very
small and subsequently arrive at the results depicted in Fig.~\ref{Fig_delta_xi_small_minus}
by tuning the parameters of asymmetries. Studying from Fig.~\ref{Fig_delta_xi_small_minus},
it is worth pointing out that this principal results are insensitive to the values of $\alpha$
and $\delta$ that measure the rotational and particle-hole asymmetries. We subsequently move
to the case with large initial values of four-fermion couplings. Paralleling the analogous
procedures in Sec.~\ref{Sec_interaction} gives rise to the same conclusion as
exhibited in Fig.~\ref{Fig_Symmetry}(b), namely the unstable of Gaussian fixed point.
Finally, we would like to investigate the situation with ``moderate initial values",
which can not trigger instability of Gaussian fixed point of fermion-fermion parameters
at the symmetric case with $\alpha=\alpha_s$ and $\delta=\delta_s$,
namely belonging to class of Fig.~\ref{Fig_Symmetry}(a).
After carrying out the similar steps, we are informed that the energy-dependent
coupling parameters $g_i$, $i=0-5$ are intimately
susceptible to the asymmetric values of $\alpha$ and $\delta$ as manifestly presented
in Fig.~\ref{Fig_delta_xi_large_minus} after paralleling the steps of previous case and
numerically evaluating coupled running Eqs.~(\ref{Eq_flow_Eq_g_0})-(\ref{Eq_flow_Eq_g_5})
with large initial values of interaction parameters $|g_i(l=0)|$. To be specific,  the four-fermion
parameters $g_i$, $i=0-5$, in analogous to Fig.~\ref{Fig_delta_xi_small_minus}, still
flow to the Gaussian fixed point if the the asymmetric values of $\alpha$ and $\delta$ are small enough as
clearly exhibited in the first column of Fig.~\ref{Fig_delta_xi_large_minus}. As long as the
asymmetric values of $\alpha$ and $\delta$ are increased, the behaviors of quartic couplings can be
completely changed. As presented in the third column of Fig.~\ref{Fig_delta_xi_large_minus},
the Gaussian fixed point is entirely destroyed by the asymmetries and the coupling parameters
$g_i$, $i=0-5$ go towards the strong coupling at the lowest-energy limit. In distinction to
the previous studies~\cite{Herbut2014PRL} where $g_0$ is stable, our results suggest that there is
no any QCP but instability towards the strong coupling caused by the contributions from the
mutual competition between the four-fermion interactions and rotational and particle-hole
asymmetries in the low-energy.

In variance with the case in the absence of the interplay between distinct
parameters, several interesting results are captured
after taking into account the fermion-fermion interactions. Before going further,
we would like to present some comments on these. To recapitulate, the system
can either go to the Gaussian fixed point or instability towards the strong coupling.
In the presence of the rotational and particle-hole symmetries, the Gaussian fixed point
would be broken and interaction parameters flow to the strong coupling by sufficiently
large values of the initial interaction strengths as exhibited in Fig.~\ref{Fig_Symmetry}(b).
Furthermore, the behaviors of four-fermion couplings are sensitive to the values
of asymmetries $\alpha$ and $\delta$ as the system does not possess the rotational
and particle-hole symmetries. While the values of parameters $\alpha$ and $\delta$ are small,
the instability hardly happens and the system always flows to
the Gaussian fixed point as depicted in the first column of Fig.~\ref{Fig_delta_xi_large_minus}.
However, the Gaussian fixed point would be destroyed totally and the system goes
towards certain instability if the values of $\alpha$ and $\delta$ are adequately
large as shown in the third column of Fig.~\ref{Fig_delta_xi_large_minus}.

\subsection{Fixed points and dominant phases}

In order to elucidate more properties of these instabilities, we are suggested to
seek the relatively fixed points at the strong coupling regime.
Commonly, the phase transitions/potential quantum phase transitions are always closely linked to
certain fixed points of interaction parameters in the low-energy regime, which are conventionally
accompanied by a multitude of singular and critical behaviors in the low-energy regime~\cite{Herbut2014PRL,Herbut2015PRB,Herbut2016PRB}. In this respect, it
is instructive to explore whether our system harbors any fixed points with the
evolutions of four-fermion parameters. To this end, we rescale all four-fermion interaction
parameters with one of them ~\cite{Vafek2012PRB,Vafek2014PRB,Wang2017}, for instance $g_5$,
to seek whether there exists any relatively fixed points in the parameter space described by
the evolutions of $g_i/g_5$, $i=0-5$ and further investigate the physical
implications for the tendency of strong couplings for the fermionic couplings.
To proceed, we derive and plot the relative trajectories for $g_i/g_5$ upon lowering
the energy scale by means of paralleling the method and analysis in last two subsections. After
carrying out several numerical calculations, we transfer the strong coupling tendencies
in Fig.~\ref{Fig_Symmetry}(b) and third column of Fig.~\ref{Fig_delta_xi_large_minus}
for both symmetric and asymmetric cases into the relative flows by rescaling all four-fermion
parameters by $g_5$ as displayed in Fig.~\ref{Fig_rescaled_FPs}. According to these results,
we are informed that the system indeed owns two kinds of relatively fixed points (RFP), namely
$(g^*_0,g^*_1,g^*_2,g^*_3,g^*_4,g^*_5)_{\mathrm{RFP-I}}\approx
(-0.054,0.017,0.027,0.022,0.022,1.00)$ for $\delta=0.3,\alpha=0.3$ and $(g^*_0,g^*_1,g^*_2,g^*_3,g^*_4,g^*_5)_{\mathrm{RFP-II}}\approx(0.180,0,0.033,0.025,
0.025,1.00)$ for $\delta=0.1,\alpha=1$.

Generally, the
interaction parameters evolving to the strong couplings or the existence of RFPs
at certain critical energy scale $l_c$ indicates the emergence of some instability
with the divergent susceptibilities of order parameters
~\cite{Vafek2012PRB,Vafek2014PRB,Maiti2010PRB,
Khodas2016PRX,Khodas2016PRB,Herbut1996PRL,Herbut1997PRL,Fu2007PRB,Ganesh2014PRL}.
To be specific, one can directly expect the emergence of superconductivity instability once
the system goes towards the RFP-II attesting to the strong couplings of attractive
fermion-fermion interactions~\cite{Shankar1994RMP,Maiti2010PRB,Herbut2016PRB}. In other words,
the superconductivity instability is linked to the dominant phase.
Next, we turn to the situation at which the RPF-I is achieved. Being different from the RFP-II,
the parameter flows $g_0$ is positively divergent upon approaching this RFP. However, we stress that
the leading phase is still associated with the superconductivity instability.
This conclusion is supported by two points. On one hand, although the parameter
$g_0$ diverges positively, its coupling $(\psi^\dagger\gamma_0\psi)^2$ conventionally
corresponds to no order~\cite{Vafek2014PRB} and is considered as a role of chemical potential.
Consequently, phase transition and dominant phase are
insusceptible to the divergence of $g_0$ no matter it is positive or negative.
On the other hand, the parameter $g_5$ that diverges negatively
is the leading one at RFP-I in that its absolute value is the largest
among all parameters. This means the attractive fermion-fermion interaction
is preferred at the low-energy limit, pointing to the superconductivity instability~\cite{Shankar1994RMP,Maiti2010PRB,Herbut2016PRB}.

\begin{figure}
\centering
\includegraphics[width=2.6in]{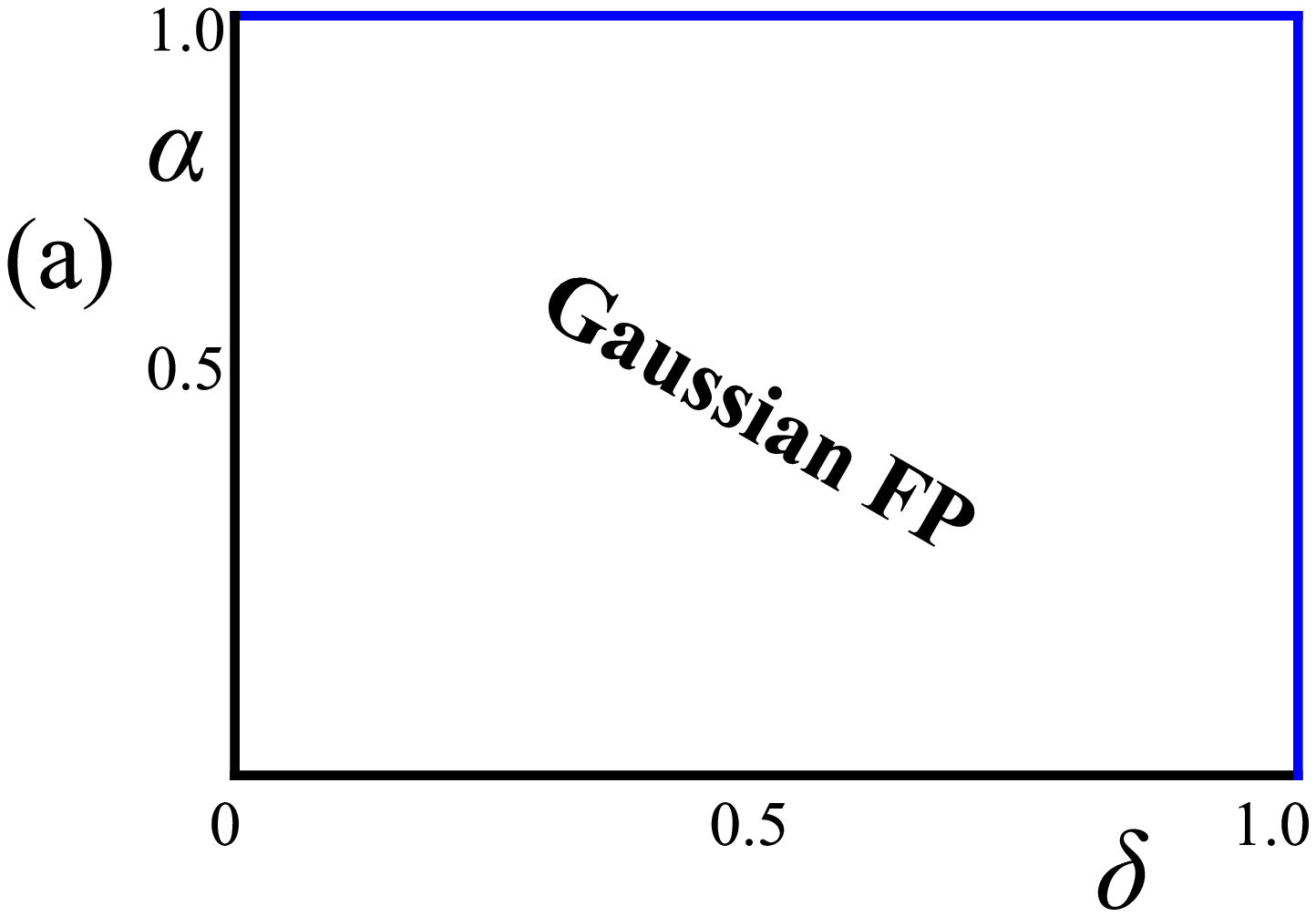}\vspace{0.2cm}
\includegraphics[width=2.6in]{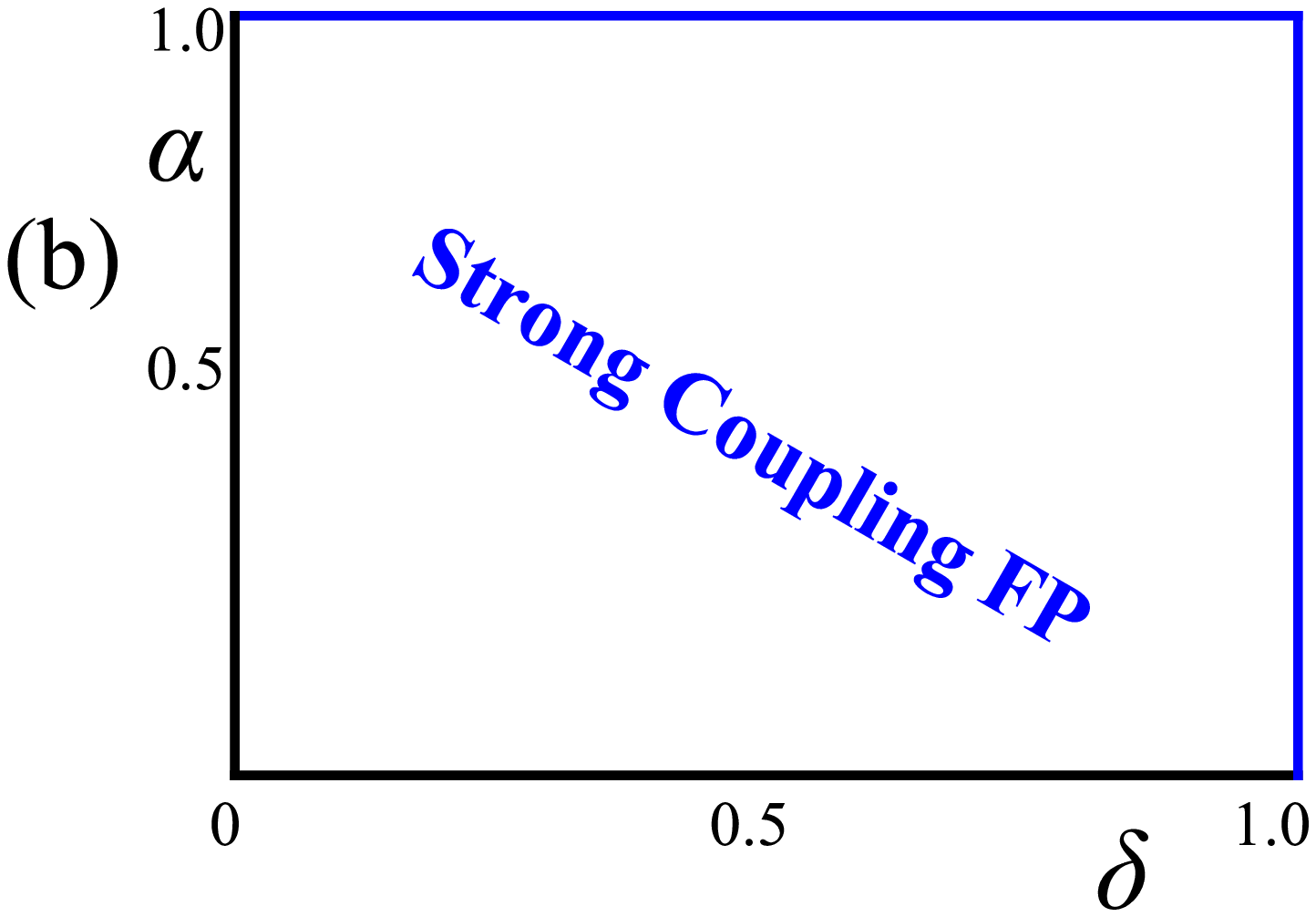}
\includegraphics[width=2.6in]{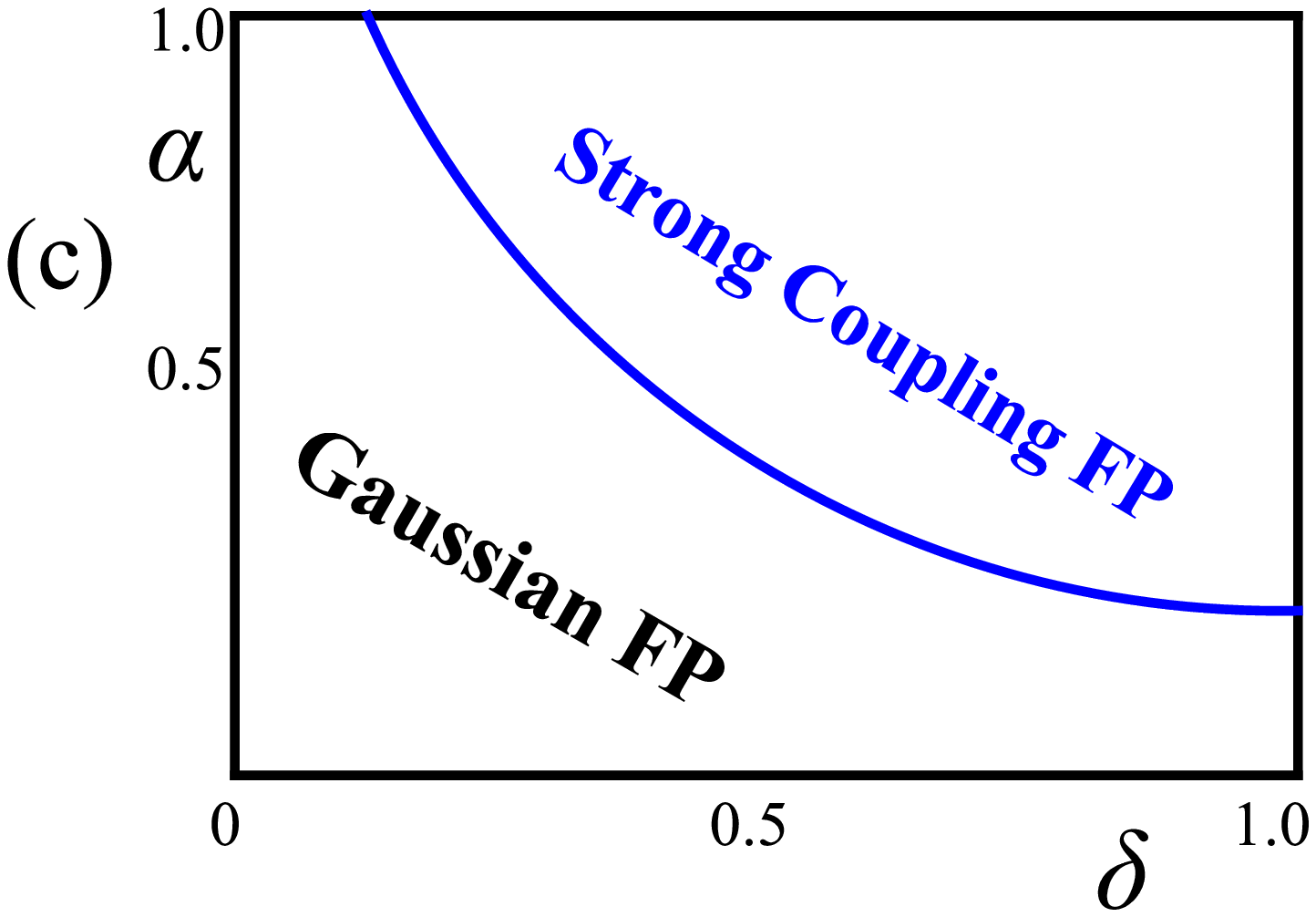}
\vspace{-0.2cm}
\caption{(Color online) The schematic phase diagrams for potential fixed points
at the lowest-energy limit in $\alpha-\delta$ plane for (a) small initial values;
(b) large initial values; and (c) moderate initial values of
interaction parameters. The strong coupling is linked to the
superconductivity instability and abbreviation ``FP" designates
the fixed point.}\label{Fig_phase_diagram_xi_delta}
\end{figure}

Based on all above analysis, we infer that the strong couplings of these four-fermion parameters
can be generated via two distinct mechanisms in the presence of fermion-fermion interactions
and asymmetries, which might be both mutually influenced and competed. On one hand, one can
strengthen the initial values of fermionic coupling parameters to induce the strong
coupling flows even at $\alpha=\alpha_s$ and $\delta=\delta_s$.
On the other, while the initial values of interaction parameters are inadequately large to
destroy the Gaussian fixed point, the rotational and particle-hole asymmetries
with $\alpha\neq\alpha_s$ and/or $\delta\neq\delta_s$ can potentially break the Gaussian fixed point
and some instability takes place with the
fermionic couplings running towards the strong coupling with the
superconductivity instability at the lowest-energy limit,.
Additionally, we find that the critical coupling strength which induces the instability
is much enhanced compared to the previous results~\cite{Herbut2014PRL}. In order to
clearly exhibit the influence of asymmetries on the physical behaviors and fixed points
and the difference between symmetric and asymmetric cases at the lowest-energy limit, we
have plotted a phase diagram as presented in Fig.~\ref{Fig_phase_diagram_xi_delta} to
overall summarize these conclusions.



%

\section{Summary}\label{Sec_summary}

In summary, we access the 3D QBT systems with certain quadratic band
touching point~\cite{Herbut2012PRB,Herbut2014PRL,Herbut2014PRB,Herbut2015PRB,
Herbut2016PRB,Herbut2017PRB,Herbut2017PRB_2}. How the low-energy
behaviors of system would be revised by the distinct sorts of fermion-fermion
interactions, rotational and particle-hole asymmetries, and their
interplay are attentively studied on the same footing by adopting
the RG approach~\cite{Shankar1994RMP,Herbut2007Book}. Not only all
six potential short-ranged fermion-fermion couplings are equally
involved, but also both symmetric and asymmetric cases are taken into
account.

The coupled flow equations of all interaction parameters
for both presence and absence of rotational and particle-hole symmetries are
explicitly addressed after performing the detailed RG analysis by collecting
the interplay between different fermionic couplings and asymmetries. Beginning
with these running equations, we can study the behaviors of
low-energy states. Initially, we find, switching on the four-fermion interactions
with rotational and particle-hole symmetries, that the QCP in noninteracting case
is destroyed and replaced by the strong couplings if the initial values of fermionic
parameters are relative large. Subsequently, we turn on the rotational and particle-hole
asymmetries. The split of the trajectories of distinct types of four-fermion couplings
are unambiguously stretched separately by the interplay between interaction and asymmetries.
In addition, certain fixed point/critical point can be induced under certain conditions,
at which the superconductivity instability is conventionally generated. To apparently
display the role of asymmetries in the physical behaviors and fixed points
as well as the difference between symmetric and asymmetric cases at the lowest-energy limit, we
have provided a schematic phase diagram provided in Fig.~\ref{Fig_phase_diagram_xi_delta}.


\section*{ACKNOWLEDGEMENTS}

J.W. is supported by the National Natural Science Foundation of China under Grant No. 11504360
and acknowledges Dr. Dmitry V. Efremov, Dr. Carmine Ortix, and Prof. Jeroen van den Brink,
for correlated collaborations and helpful discussions as well as Prof. W. Liu
for useful discussions.

\end{document}